# Robust and Efficient Distributed Compression for Cloud Radio Access Networks

Seok-Hwan Park, Osvaldo Simeone, Onur Sahin and Shlomo Shamai (Shitz)

## Abstract

This work studies distributed compression for the uplink of a cloud radio access network where multiple multi-antenna base stations (BSs) are connected to a central unit, also referred to as cloud decoder, via capacity-constrained backhaul links. Since the signals received at different BSs are correlated, distributed source coding strategies are potentially beneficial, and can be implemented via sequential source coding with side information. For the problem of compression with side information, available compression strategies based on the criteria of maximizing the achievable rate or minimizing the mean square error are reviewed first. It is observed that, in either case, each BS requires information about a specific covariance matrix in order to realize the advantage of distributed source coding. Since this covariance matrix depends on the channel realizations corresponding to other BSs, a robust compression method is proposed for a practical scenario in which the information about the covariance available at each BS is imperfect. The problem is formulated using a deterministic worst-case approach, and an algorithm is proposed that achieves a stationary point for the problem. Then, BS selection is addressed with the aim of reducing the number of active BSs, thus enhancing the energy efficiency of the network. An optimization problem is formulated in which compression and BS selection are performed jointly by introducing a sparsity-inducing term into the objective function. An iterative algorithm is proposed that is shown to converge to a locally optimal point. From numerical results, it is observed that the proposed robust compression scheme compensates for a large fraction of the performance loss induced by the imperfect statistical information. Moreover, the proposed BS selection algorithm is seen to perform close to the more complex exhaustive search solution.

S.-H. Park and O. Simeone are with the Center for Wireless Communications and Signal Processing Research (CWCSPR), ECE Department, New Jersey Institute of Technology (NJIT), Newark, NJ 07102, USA (email: {seok-hwan.park, osvaldo.simeone}@njit.edu).

O. Sahin is with InterDigital Inc., Melville, New York, 11747, USA (email: Onur.Sahin@interdigital.com).

S. Shamai (Shitz) is with the Department of Electrical Engineering, Technion, Haifa, 32000, Israel (email: sshlomo@ee.technion.ac.il).

# I. INTRODUCTION

The current deployments of cellular systems are facing the "bandwidth crunch" problem caused by the ever increasing demand for high data rate applications. An integral part of many proposed solutions to this problem is the idea of cloud radio access networks, whereby the baseband processing of the base stations (BSs) is migrated to a central unit in the "cloud" to which the BSs are connected via backhaul links (see, e.g., [1]). This enables an effective implementation of the principle of network MIMO [2] that has the further advantages of simplifying the deployment and management of BSs and of reducing BS energy consumption. We refer to [2] for a thorough review of previous work on network MIMO.

On the uplink of a cloud radio access network, the BSs operate as terminals that relay "soft" information to the cloud decoder regarding the received baseband signals (see Fig. 1). Since the signals received at different BSs are correlated, distributed source coding strategies are generally beneficial, as first demonstrated in [3]-[5]. In this paper, we study the problem of compression with distributed source coding in the presence of multi-antenna BSs by focusing on the issues of *robustness* and *efficiency*. Specifically, the performance of distributed source coding is known to be sensitive to imperfections in the knowledge of the joint statistics of the received signals to be compressed. We tackle this problem by formulating *robust* version of distributed source coding. Moreover, an efficient operation of cloud radio access networks is understood to require a parsimonious use of the BSs, whose energy consumption is among the most relevant contributions to the overall energy expenditure for the network. We address this problem by proposing a joint compression and BS selection approach.

## A. Contributions and Related Work

Related works on uplink multi-cell processing with constrained backhaul can be found in [5]-[8] and references therein. In [5], compress-and-forward strategies with joint decompression and decoding were proposed for the uplink of a Wyner model with single-antenna terminals. The set-up with multiple antennas but a single transmitter is studied in [4]. Various relaying schemes, including decode-and-forward and compress-and-forward techniques, were compared in [6]. In [7], a different relaying scheme was proposed, which is referred to as compute-and-forward, in which structured codes are used by Mobile Stations (MSs), and the BSs decode a function of the transmitted messages. An efficient implementation of the compute-and-forward scheme, which is referred to as quantized compute-and-forward, was proposed in [8].

In this work, we focus on the distributed compression of the received signals at multiple multi-antenna BSs. Distributed compression can be implemented via sequential source coding with side information [9]: at each step of this process, a BS compresses its received signal by leveraging the available statistical information about the signals compressed by the BSs active at the previous steps[1]. The problem of compressing Gaussian random vector signals (as with a multi-antenna receiver in the presence of Gaussian codebooks and Gaussian noise) with receiver side information has been investigated in [10]-[14]. Specifically, it was shown in [10] and [11] that independent coding of the signals obtained at the output of the so called conditional Karhunen-Loeve transform (KLT) achieves the optimal performance in terms of minimizing the mean square error (MMSE) and maximizing achievable rate (Max-Rate), respectively.

In this work, we first provide a review of the MMSE and Max-Rate [10][11] compression strategies for the uplink of a cloud radio access cellular network that employs sequential source coding with side information. In the process, we will also point out the connection with the so called *information bottleneck problem* [12][13][15] and some consequence of this observation (Sec. III).

The performance of distributed source coding is very sensitive to errors in the knowledge of the joint statistics of the received signals at the BSs due to the potential inability of the cloud decoder to decompress the signal received by a BS. This is because distributed source coding is based on the idea of reducing the rate of the compressed stream by introducing some uncertainty on the compressed signal that is resolved with the aid of the side information [16]. The amount of rate reduction that is allowed without incurring decompression errors thus depends critically on the quality of the side information, which should be known to the encoder.

Motivated by this observation, in Sec. IV, we propose a *robust compression scheme* by assuming the knowledge of the joint statistics, which amount here to a covariance matrix, available at each BS is imperfect. To model the uncertainty, we adopt a deterministic additive error model with bounds on eigenvalues of the error matrix similar to [17]-[19] (see also [20]).

---

[1]This argument assumes, as done throughout this paper, that the cloud decoder performs decompression of the received signals and decoding separately. For analysis of joint decompression and decoding for single-antenna BSs, we refer to [3].

We remark that bounding the eigenvalues is equivalent to bounding any norm of the error [21, Appendix A]. The problem is formulated following a deterministic worst-case approach and a solution that achieves a stationary point of this problem is provided by solving Karush-Kuhn-Tucker (KKT) conditions [21][22], which are also shown to be necessary for optimality.

We then tackle the issue of *energy efficient network operation* via selection of the the BSs. This problem was tackled for a multi-antenna downlink system in [23] and for a single-antenna uplink system in [24]. In Sec. V, an optimization problem is formulated in which compression and BS selection are performed jointly by introducing a sparsity-inducing term into the objective function. This follows the strategy proposed in [23] for the design of beamforming vectors for the downlink. An iterative block-coordinate ascent algorithm is proposed that is shown to converge to a locally optimal point. We conclude the paper with numerical results in Sec. VI.

*Notation*: We use the same notation for probability mass functions (pmfs) and probability density functions (pdfs), namely $p(x)$ represents the distribution, pmf or pdf, of random variable $X$. Similar notations are used for joint and conditional distributions. All logarithms are in base two unless specified. Given a vector $\mathbf{x} = [x_1, x_2.., x_n]^T$, we define $\mathbf{x}_{\mathcal{S}}$ for a subset $\mathcal{S} \subseteq \{1, 2, ..., n\}$ as the vector including, in arbitrary order, the entries $x_i$ with $i \in \mathcal{S}$. Notation $\mathbf{\Sigma_x}$ is used for the correlation matrix of random vector $\mathbf{x}$, i.e., $\mathbf{\Sigma_x} = \mathrm{E}[\mathbf{xx}^\dagger]$; $\mathbf{\Sigma_{xy}}$ represents the cross-correlation matrix $\mathbf{\Sigma_{xy}} = \mathrm{E}[\mathbf{xy}^\dagger]$; and $\mathbf{\Sigma_{x|y}}$ represents the "conditional" correlation matrix of $\mathbf{x}$ given $\mathbf{y}$, namely $\mathbf{\Sigma_{x|y}} = \mathbf{\Sigma_x} - \mathbf{\Sigma_{xy}} \mathbf{\Sigma_y^{-1}} \mathbf{\Sigma_{xy}^\dagger}$. Notation $\mathcal{H}^n$ represents the set of all $n \times n$ Hermitian matrices.

## II. System Model

We consider a *cluster* of cells[2], which includes a total number $N_B$ of BSs, each being either an MBS or a HBS, and there are $N_M$ active MSs. We refer to Fig. 1 for an illustration. We denote the set of all BSs as $\mathcal{N}_B = \{1, ..., N_B\}$. Each $i$th BS is connected to the cloud decoder via a finite-capacity link of capacity $C_i$ and has $n_{B,i}$ antennas, while each MS has $n_{M,i}$ antennas. Throughout the paper, we focus on the uplink.

Defining $\mathbf{H}_{ij}$ as the $n_{B,i} \times n_{M,j}$ channel matrix between the $j$th MS and the $i$th BS, the overall

---

[2]The model applies also to a cluster of sectors.

 

channel matrix toward BS $i$ is given as the $n_{B,i} \times n_M$ matrix

$$\mathbf{H}_i = [\mathbf{H}_{i1} \cdots \mathbf{H}_{iN_M}], \tag{1}$$

with $n_M = \sum_{i=1}^{N_M} n_{M,i}$. Assuming that all the $N_M$ MSs in a cluster are synchronous, at any discrete-time channel use (c.u.) of a given time-slot, the signal received by the $i$th BS is given by

$$\mathbf{y}_i = \mathbf{H}_i \mathbf{x} + \mathbf{z}_i. \tag{2}$$

In (2), vector $\mathbf{x} = [\mathbf{x}_1^\dagger \cdots \mathbf{x}_{N_M}^\dagger]^\dagger$ is the $n_M \times 1$ vector of symbols transmitted by all the MSs in the cluster at hand. The noise vectors $\mathbf{z}_i$ are independent over $i$ and are distributed as $\mathbf{z}_i \sim \mathcal{CN}(0, \mathbf{I})$, for $i \in \{1, ..., N_B\}$. Note that the noise covariance matrix is selected as the identity without loss of generality, since the received signal can always be whitened by the BSs. The channel matrix $\mathbf{H}_i$ is assumed to be constant in each time-slot and, unless stated otherwise, is considered to be known at the cloud decoder.

Using standard random coding arguments, the coding strategies employed by the MSs in each time-slot entail a distribution $p(\mathbf{x})$ on the transmitted signals that factorizes as

$$p(\mathbf{x}) = \prod_{i=1}^{N_M} p(\mathbf{x}_i), \tag{3}$$

since the signals sent by different MSs are independent. Note that the signals $\mathbf{x}$ are typically discrete, e.g., taken from discrete constellation, but can be well approximated by continuous (e.g., Gaussian) distributions for capacity-achieving codes over Gaussian channels. If not stated otherwise, we will thus assume throughout that the distribution $p(\mathbf{x}_i)$ of the signal transmitted by the $i$th MS is given as $\mathbf{x}_i \sim \mathcal{CN}(0, \mathbf{\Sigma}_{\mathbf{x}_i})$ for a given covariance matrix $\mathbf{\Sigma}_{\mathbf{x}_i}$.

The BSs communicate with the cloud by providing the latter with *soft information* derived from the received signal. Note that the BSs do not need to be informed about the MSs' codebooks [5]. Using a conventional rate-distortion theory arguments, a compression strategy for the $i$th BS is described by a test channel $p(\hat{\mathbf{y}}_i|\mathbf{y}_i)$ that describes the relationship between the signal to be compressed, namely $\mathbf{y}_i$, and its description $\hat{\mathbf{y}}_i$ to be communicated to the cloud (see, e.g., [25]). It is recalled that such compression is limited to $C_i$ bits per received symbol. The cloud decodes jointly the signals $\mathbf{x}$ of all MSs based on all the descriptions $\hat{\mathbf{y}}_i$ for $i \in \mathcal{N}_\mathcal{B}$, so that, from standard information-theoretic considerations, the achievable sum-rate is given by

$$R_{sum} = I(\mathbf{x}; \hat{\mathbf{y}}_{\mathcal{N}_\mathcal{B}}). \tag{4}$$

Since the signals $\mathbf{y}_i$ measured by different BSs are correlated, distributed source coding techniques have the potential to improve the quality of the descriptions $\hat{\mathbf{y}}_i$ [3]. Specifically, given compression test channels $p(\hat{\mathbf{y}}_i|\mathbf{y}_i)$, it is well known that the descriptions $\hat{\mathbf{y}}_i$ can be recovered at the cloud as long as the capacities $C_i$ satisfy the following conditions (see, e.g., [25, Ch. 11])

$$\sum_{j \in \mathcal{S}} C_j \geq I(\mathbf{y}_{\mathcal{S}}; \hat{\mathbf{y}}_{\mathcal{S}}|\hat{\mathbf{y}}_{\bar{\mathcal{S}}}) \text{ for all } \mathcal{S} \subseteq \mathcal{N}_B \tag{5}$$

where we have defined $\bar{\mathcal{S}} = \mathcal{N}_B - \mathcal{S}$. We recall that $\mathbf{y}_{\mathcal{S}}$ includes all $\mathbf{y}_i$ with $i \in \mathcal{S}$ and similarly for $\hat{\mathbf{y}}_{\mathcal{S}}$ and $\hat{\mathbf{y}}_{\bar{\mathcal{S}}}$.

We now recall a known property of the region of backhaul capacities described by (5).

**Lemma 1.** [9] *The polytope[3] described by the inequalities (5) (along with $C_i \geq 0$ for $i \in \mathcal{N}_B$) in the space of the capacities $C_i$ is the convex hull of $N_B!$ vertices, one for each permutation $\pi(i)$ of the indices $i \in \{1, ..., N_B\}$[4]. Specifically, the vertex corresponding to the $i$th element of a permutation $\pi(i)$ is given by*

$$C_{\pi(i)} = I(\mathbf{y}_{\pi(i)}; \hat{\mathbf{y}}_{\pi(i)}|\hat{\mathbf{y}}_{\{\pi(1),...,\pi(i-1)\}}) \text{ for all } i = 1, ..., N_B. \tag{6}$$

*Remark* 1. In the approach discussed above, the central processor in the cloud decodes the descriptions $\hat{\mathbf{y}}_{\mathcal{N}_B}$ received from the BSs and then perform joint decoding of all of the MSs' signals. An alternative approach, discussed in [3], is instead that of performing joint decoding of both the descriptions $\hat{\mathbf{y}}_{\mathcal{N}_B}$ and of the signals $\mathbf{x}$ transmitted by the MSs. We will not further consider this approach here (see also Sec. VII).

*Remark* 2. The Gaussian distribution assumed here for the transmitted signals $\mathbf{x}$ maximizes the capacity of a Gaussian channel, but is not in general optimal for the setting at hand in which the receiver observes a compressed version of the received signal, even in a system with only one BS ($N_B = 1$) [3].

## III. Maximizing the Sum-Rate

In this section, we first discuss a greedy approach to find a suboptimal solution to the problem of maximizing the sum-rate (4). Then, we will focus on the main step of this greedy procedure

---

[3]More specifically, the polytope at hand is a contra-polymatroid.

[4]A permutation is a function $\pi(i)$ from set $\{1, ..., N_B\}$ to the same set such that $\pi(i) \neq \pi(j)$ for all $i \neq j$.

by reviewing known results and pointing out some new observation along the way.

## A. Problem Definition and Greedy Solution

In this section, we aim at maximizing the sum-rate (4) under the constraints (5). If we restrict the search only to the vertices of the rate region (5), we can obtain a generally suboptimal[5] solution by solving the problem

$$\underset{\pi, \{p(\hat{\mathbf{y}}_i | \mathbf{y}_i)\}_{i \in \mathcal{N}_B}}{\text{maximize}} \quad I(\mathbf{x}; \hat{\mathbf{y}}_{\mathcal{N}_B}) \tag{7}$$

where the optimization is subject to the constraints (6) and the optimization space includes also the BS permutation $\pi$.

The optimization problem (7) is generally still complex. We thus propose a greedy approach in Algorithm 1 to the selection of the permutation $\pi$, while optimizing the test channels $p(\hat{\mathbf{y}}_i | \mathbf{y}_i)$ at each step of the greedy algorithm. The greedy algorithm is based on the chain rule for the mutual information, that allows the sum-rate (4) to be written as

$$I(\mathbf{x}; \hat{\mathbf{y}}_{\mathcal{N}_B}) = \sum_{i=1}^{N_B} I(\mathbf{x}; \hat{\mathbf{y}}_{\pi(i)} | \hat{\mathbf{y}}_{\{\pi(1), \dots, \pi(i-1)\}}) \tag{8}$$

for any permutation $\pi$ of the set $\{1, \dots, N_B\}$. As a result of the algorithm, we obtain a permutation $\pi^*$ and feasible (in the sense of satisfying constraint (5)) test channels $p^*(\hat{\mathbf{y}}_i | \mathbf{y}_i)$. From now on, we refer to the compression based on (9) in Algorithm 1 as Max-Rate compression.

*Remark* 3. The implementation of the greedy algorithm in Algorithm 1 requires solving the problem (9) for each $i$th BS for a given order $\pi$. In practice, problem (9) can be solved at the cloud decoder, which then communicates the result to the $i$th BS. As it will be further clarified below, this approach requires the cloud center to know the channel matrices $\mathbf{H}_i$, $i = 1, \dots, N_B$. Alternatively, once the order $\pi$ is fixed by the cloud, problem (9) can be solved at each $i$th BS. This second approach requires the cloud to communicate some information to each BS, as it will be seen in the following.

---

[5]While it is easy to see that the optimum of the problem above is achieved at the boundary of the rate region (5), it is not clear whether the optimum point is in general one of the vertices.

---

**Algorithm 1** Greedy Approach to the selection of the ordering $\pi$ and the test channels $p(\hat{\mathbf{y}}_i|\mathbf{y}_i)$

1. Initialize set $\mathcal{S}$ to be an empty set, i.e., $\mathcal{S}^{(0)} = \varnothing$ .

2. For $j = 1, \ldots, N_B$, perform the following steps.

   i) Each $i$th BS with $i \in \mathcal{N}_B - \mathcal{S}$ evaluates the test channel $p(\hat{\mathbf{y}}_i|\mathbf{y}_i)$ by solving the problem

$$\underset{p(\hat{\mathbf{y}}_i|\mathbf{y}_i)}{\text{maximize}} \; I(\mathbf{x}; \hat{\mathbf{y}}_i|\hat{\mathbf{y}}_{\mathcal{S}})$$

$$\text{s.t. } I(\mathbf{y}_i; \hat{\mathbf{y}}_i|\hat{\mathbf{y}}_{\mathcal{S}}) \leq C_i. \tag{9}$$

   Denote the optimal value of this problem as $\phi_i^\star$ and an optimal test channel as $p^*(\hat{\mathbf{y}}_i|\mathbf{y}_i)$.

   ii) Choose the BS $i \in \mathcal{N}_B - \mathcal{S}$ with the largest optimal value $\phi_i^\star$ and add it to the set $\mathcal{S}$,

    i.e., $\mathcal{S}^{(j)} = \mathcal{S}^{(j-1)} \cup \{i\}$ and set $\pi^*(j) = i$.

---

### B. Max-Rate Compression

We now discuss the solution to the problem (9) of optimizing the compression test channel $p(\hat{\mathbf{y}}_i|\mathbf{y}_i)$ at the $i$th BS under the assumption that the cloud decoder has side information $\hat{\mathbf{y}}_{\mathcal{S}}$ with $\mathcal{S} = \{\pi(1), \ldots, \pi(i-1)\}$. Note that the random vectors involved in problem (9) satisfy the Markov chain $\hat{\mathbf{y}}_{\mathcal{S}} \leftrightarrow \mathbf{x} \leftrightarrow \mathbf{y}_i \leftrightarrow \hat{\mathbf{y}}_i$. We first review the solution of problem (9) given in [11]. We also point out the relationship of the solution found in [11] with the information bottleneck method for Gaussian variables of reference [13]. This connection does not seem to have been observed before and allows for a solution of problem (9) in the presence of a generic discrete distribution (3) of the transmitted signals, as briefly discussed in Remark 5. We then briefly review more conventional compression techniques based on MMSE compression.

We first observe that, under the given assumption of Gaussian input vector $\mathbf{x}$, the optimal test channel $p(\hat{\mathbf{y}}_i|\mathbf{y}_i)$ for problem (9) is given by a Gaussian distribution as stated in the following lemma.

**Lemma 2.** [11][13] *For a Gaussian distributed* $\mathbf{x}$*, the optimal test channel* $p(\hat{\mathbf{y}}_i|\mathbf{y}_i)$ *is characterized as*

$$\hat{\mathbf{y}}_i = \mathbf{A}_i \mathbf{y}_i + \mathbf{q}_i \tag{10}$$

*where* $\mathbf{A}_i$ *is a matrix to be calculated and* $\mathbf{q}_i \sim \mathcal{CN}(0, \mathbf{I})$ *is the compression noise, which is* *independent of* $\mathbf{x}$ *and* $\mathbf{z}_i$. *Moreover, defining* $\boldsymbol{\Omega}_i = \mathbf{A}_i^\dagger \mathbf{A}_i$, *problem (9) can be restated as*

$$\underset{\boldsymbol{\Omega}_i \succeq \mathbf{0}}{\text{maximize}} \; f_i\left(\boldsymbol{\Omega}_i\right) - \log \det \left(\mathbf{I} + \boldsymbol{\Omega}_i\right) \tag{11}$$

$$\text{s.t.} \quad f_i\left(\boldsymbol{\Omega}_i\right) \leq C_i,$$

*where the function* $f_i\left(\boldsymbol{\Omega}_i\right)$ *is defined as*

$$f_i\left(\boldsymbol{\Omega}_i\right) = \log \det \left(\mathbf{I} + \boldsymbol{\Omega}_i \left(\mathbf{H}_i \boldsymbol{\Sigma}_{\mathbf{x}|\hat{\mathbf{y}}_S} \mathbf{H}_i^\dagger + \mathbf{I}\right)\right), \tag{12}$$

*and we have*

$$\boldsymbol{\Sigma}_{\mathbf{x}|\hat{\mathbf{y}}_S} = \boldsymbol{\Sigma}_{\mathbf{x}} - \boldsymbol{\Sigma}_{\mathbf{x}} \bar{\mathbf{H}}_S^\dagger \left(\bar{\mathbf{H}}_S \boldsymbol{\Sigma}_{\mathbf{x}} \bar{\mathbf{H}}_S^\dagger + \boldsymbol{\Sigma}_{\mathbf{t}_S}\right)^{-1} \bar{\mathbf{H}}_S \boldsymbol{\Sigma}_{\mathbf{x}}. \tag{13}$$

*Proof:* We refer readers to [11][13] for the proof. ∎

With compression model (10), at the $i$th stage in the greedy algorithm, the side information $\hat{\mathbf{y}}_S$ available to the cloud decoder is given as

$$\hat{\mathbf{y}}_{\pi(j)} = \bar{\mathbf{H}}_{\pi(j)} \mathbf{x} + \mathbf{t}_{\pi(j)}, \tag{14}$$

where $\bar{\mathbf{H}}_{\pi(j)} = \mathbf{A}_{\pi(j)} \mathbf{H}_{\pi(j)}$ and $\mathbf{t}_{\pi(j)} = \mathbf{A}_{\pi(j)} \mathbf{z}_{\pi(j)} + \mathbf{q}_{\pi(j)}$ for $j = 1, \cdots, i-1$.

**Proposition 1.** [11] An optimal solution $\boldsymbol{\Omega}_i^*$ to problem (11) is given by

$$\boldsymbol{\Omega}_i^* = \mathbf{U} \text{diag} \left(\alpha_1, \ldots, \alpha_{n_{B,i}}\right) \mathbf{U}^\dagger, \tag{15}$$

where we have the eigenvalue decomposition $\mathbf{H}_i \boldsymbol{\Sigma}_{\mathbf{x}|\hat{\mathbf{y}}_S} \mathbf{H}_i^\dagger + \mathbf{I} = \mathbf{U} \text{diag} \left(\lambda_1, \ldots, \lambda_{n_{B,i}}\right) \mathbf{U}^\dagger$ with unitary $\mathbf{U}$ and ordered eigenvalues $\lambda_1 \geq \cdots \geq \lambda_{n_{B,i}}$. The diagonal elements $\alpha_1, \ldots, \alpha_{n_{B,i}}$ are computed as

$$\alpha_l = \left[\frac{1}{\mu}\left(1 - \frac{1}{\lambda_l}\right) - 1\right]^+, \; l = 1, \ldots, n_{B,i}, \tag{16}$$

where $\mu$ is such that the condition $\sum_{l=1}^{n_{B,i}} \log\left(1 + \alpha_l \lambda_l\right) = C_i$ is satisfied.

*Remark* 4. The linear transformation $\mathbf{U}$ is referred to as a conditional KLT in [10]. An illustration of the optimal Max-Rate compression is shown in Fig. 2.

*Remark* 5. An alternative formulation of the optimal solution of Proposition 1 can be obtained using the results in [12, Theorem 3.1]. In fact, problem (9) can be interpreted as an instance of the information bottleneck problem, which was introduced in [15]. We recall that the information bottleneck method consists in the maximization of $I(\mathbf{x}; \hat{\mathbf{y}}) - 1/\beta I(\mathbf{y}; \hat{\mathbf{y}})$ for a given $\beta > 0$ over the conditional distribution $p(\hat{\mathbf{y}}|\mathbf{y})$ for random variables $\mathbf{x}, \mathbf{y}, \hat{\mathbf{y}}$ satisfying the Markov chain $\mathbf{x} \leftrightarrow \mathbf{y} \leftrightarrow \hat{\mathbf{y}}$. This connection between the information bottleneck problem and that of compression for cloud radio access network allows us to import tools developed for the information bottleneck problem to the set-up at hand [15]. For instance, to solve problem (9) in the case where the distribution of the transmitted symbols (3) is not Gaussian but discrete, it may be advantageous to use a discrete alphabet for $\hat{\mathbf{y}}_i$. Assuming this alphabet to be given, the information bottleneck approach enables us to maximize $I(\mathbf{x}; \hat{\mathbf{y}}_i | \hat{\mathbf{y}}_{\mathcal{S}}) - 1/\beta I(\mathbf{y}_i; \hat{\mathbf{y}}_i | \hat{\mathbf{y}}_{\mathcal{S}})$, or equivalently to minimize $I(\mathbf{y}_i; \hat{\mathbf{y}}_i | \hat{\mathbf{y}}_{\mathcal{S}}) - \beta I(\mathbf{x}; \hat{\mathbf{y}}_i | \hat{\mathbf{y}}_{\mathcal{S}})$, over the pmf $p(\hat{\mathbf{y}}_i|\mathbf{y}_i)$ for some Lagrange multiplier $\beta > 0$. To this end, an iterative algorithm from [15, Sec. 3.3] can be adopted with minor modifications, which are due to the conditioning on $\hat{\mathbf{y}}_{\mathcal{S}}$ that does not appear in [15]. We recall that the key idea of the algorithm is to maximize the objective function over the three distributions $p(\hat{\mathbf{y}}_i|\hat{\mathbf{y}}_{\mathcal{S}})$, $p(\hat{\mathbf{y}}_i|\hat{\mathbf{y}}_i)$ and $p(\mathbf{x}|\hat{\mathbf{y}}_i, \hat{\mathbf{y}}_{\mathcal{S}})$ in turn. Since the objective function can be seen to be a concave function in the domain of the three distributions, an iterative block-coordinate ascent algorithm is known to converge to a locally optimal solution [22]. We do not further detail this approach here.

## C. MMSE Compression

Here we review conventional approaches that are, however, suboptimal for problem (9). Specifically, we review techniques that aim at minimizing the MSE, which we refer to as Minimum MSE (MMSE) techniques.

**Direct MMSE, No Side Information (NSI):** In the most conventional approach, compression is done with the aim of minimizing the MSE on the received signal $\mathbf{y}_i$, and neglecting the fact that the cloud decoder has side information $\hat{\mathbf{y}}_{\mathcal{S}}$. This leads to the following compression criterion:

$$\underset{p(\hat{\mathbf{y}}_i|\mathbf{y}_i), g(\hat{\mathbf{y}}_i)}{\text{minimize}} \ \mathrm{E}[||g(\hat{\mathbf{y}}_i) - \mathbf{y}_i||^2] \tag{17}$$

$$\text{s.t.} \ I(\mathbf{y}_i; \hat{\mathbf{y}}_i) \leq C_i,$$

where $g(\cdot)$ is some function (such as MMSE estimate) that can be affected at the decoder's side. We note that the constraint in problem (17) involves a mutual information that is not conditioned on $\hat{\mathbf{y}}_{\mathcal{S}}$, as the side information at the cloud is not leveraged here [25].

**Indirect MMSE, No Side Information (NSI):** A potentially better compression can be obtained by accounting for the fact that the cloud is interested in recovering $\mathbf{x}$ and not $\mathbf{y}_i$, and thus using the MMSE criterion

$$\underset{p(\hat{\mathbf{y}}_i|\mathbf{y}_i), g(\hat{\mathbf{y}}_i)}{\text{minimize}} \quad \mathrm{E}[||g(\hat{\mathbf{y}}_i) - \mathbf{x}||^2] \tag{18}$$

$$\text{s.t. } I(\mathbf{y}_i; \hat{\mathbf{y}}_i) \le C_i,$$

where function $g(\cdot)$ has the same interpretation as above. We refer to this strategy as "indirect" MMSE since it compresses $\mathbf{y}_i$ but with the goal of providing a description of $\mathbf{x}$, which is not directly measured at the BSs (this nomenclature is standard, see, e.g., [26]).

By leveraging the side information $\hat{\mathbf{y}}_{\mathcal{S}}$, which is available at the cloud decoder, we can obtain improved versions of the direct and indirect MMSE strategies discussed above. The side information $\hat{\mathbf{y}}_{\mathcal{S}}$ can be used for two purposes: (*i*) to reduce the compression rate from $I(\mathbf{y}_i; \hat{\mathbf{y}}_i)$ to $I(\mathbf{y}_i; \hat{\mathbf{y}}_i|\hat{\mathbf{y}}_{\mathcal{S}})$; (*ii*) to obtain the final estimate at the cloud decoder as a function $g(\hat{\mathbf{y}}_i, \hat{\mathbf{y}}_{\mathcal{S}})$ of both the description $\hat{\mathbf{y}}_i$ and the side information $\hat{\mathbf{y}}_{\mathcal{S}}$ (see, e.g., [25]).

**Direct MMSE, Side Information (SI):** The direct MMSE method that leverages side information leads to the criterion

$$\underset{p(\hat{\mathbf{y}}_i|\mathbf{y}_i),\, g(\hat{\mathbf{y}}_i, \mathbf{y}_{\mathcal{S}})}{\text{minimize}} \quad \mathrm{E}[||g(\hat{\mathbf{y}}_i, \hat{\mathbf{y}}_{\mathcal{S}}) - \mathbf{y}_i||^2] \tag{19}$$

$$\text{s.t. } I(\mathbf{y}_i; \hat{\mathbf{y}}_i|\hat{\mathbf{y}}_{\mathcal{S}}) \le C_i.$$

**Indirect MMSE, Side Information (SI):** Similarly, the indirect MMSE method instead leads to the criterion

$$\underset{p(\hat{\mathbf{y}}_i|\mathbf{y}_i),\, g(\hat{\mathbf{y}}_i, \mathbf{y}_{\mathcal{S}})}{\text{minimize}} \quad \mathrm{E}[||g(\hat{\mathbf{y}}_i, \hat{\mathbf{y}}_{\mathcal{S}}) - \mathbf{x}||^2] \tag{20}$$

$$\text{s.t. } I(\mathbf{y}_i; \hat{\mathbf{y}}_i|\hat{\mathbf{y}}_{\mathcal{S}}) \le C_i.$$

We now review the solution to these problems. The solutions are obtained by leveraging the fairly well known results (see, in particular, [10][14] for recent treatments).

**Proposition 2.** The optimal compression for the MMSE criteria listed above is given as (10) with $\mathbf{A}_i$ such that $\mathbf{\Omega}_i = \mathbf{A}_i^\dagger \mathbf{A}_i$ and

$$\mathbf{\Omega}_i = \mathbf{P}_i^\dagger \mathbf{U} \text{diag} \left( \alpha_1, \ldots, \alpha_{n_{B,i}} \right) \mathbf{U}^\dagger \mathbf{P}_i, \tag{21}$$

where $\mathbf{P}_i = \mathbf{I}$ for direct methods and $\mathbf{P}_i = \mathbf{\Sigma}_{\mathbf{x}\mathbf{y}_i} \mathbf{\Sigma}_{\mathbf{y}_i}^{-1}$ for indirect methods (where $\mathbf{\Sigma}_{\mathbf{x}\mathbf{y}_i} = \mathbf{\Sigma}_{\mathbf{x}} \mathbf{H}_i^\dagger$ and $\mathbf{\Sigma}_{\mathbf{y}_i} = \mathbf{H}_i \mathbf{\Sigma}_{\mathbf{x}} \mathbf{H}_i^\dagger + \mathbf{I}$). The columns of matrix $\mathbf{U}$ are the eigenvectors of covariance matrices $\Sigma_{\bar{\mathbf{y}}_i} = \mathbf{P}_i \left( \mathbf{H}_i \Sigma_{\mathbf{x}} \mathbf{H}_i^\dagger + \mathbf{I} \right) \mathbf{P}_i^\dagger$ and $\Sigma_{\bar{\mathbf{y}}_i | \hat{\mathbf{y}}_S} = \mathbf{P}_i \left( \mathbf{H}_i \Sigma_{\mathbf{x} | \hat{\mathbf{y}}_S} \mathbf{H}_i^\dagger + \mathbf{I} \right) \mathbf{P}_i^\dagger$ for the NSI and SI cases, respectively, corresponding to the ordered eigenvalues $\lambda_1 \geq ... \geq \lambda_{n_{B,i}}$, and

$$\alpha_l = \left[ \frac{1}{\mu} - \frac{1}{\lambda_l} \right]^+ \tag{22}$$

where $\mu$ is such that the condition $\sum_{l=1}^{n_{B,i}} \log\left( 1 + \alpha_l \lambda_l \right) = C_i$ is satisfied.

*Proof:* The proof of Proposition 2 follows immediately by conditional KLT results of [10, Sec. III]. ∎

*Remark 6.* In Proposition 2, matrix $\mathbf{P}_i = \Sigma_{\mathbf{x}\mathbf{y}_i} \Sigma_{\mathbf{y}_i}^{-1}$ is used in "indirect MMSE" methods for pre-processing and gives the MMSE estimate $\bar{\mathbf{y}}_i$ of $\mathbf{x}$ given $\mathbf{y}_i$. This pre-processing equation transforms the indirect compression problem into a direct one in which the signal to be compressed according to the MMSE criterion is $\bar{\mathbf{y}}_i = \mathbf{P}_i \mathbf{y}_i$ (see, e.g., [14]).

*Remark 7.* By comparing Proposition 1 and Proposition 2, it can be seen that the direct MMSE method with SI and the Max-Rate solution differ only in the way the given $\alpha_1, \ldots, \alpha_{n_{B,i}}$ in (15) are chosen. In fact, it can be seen that the MMSE gain allocation (22) over the streams after the KLT depends only on the corresponding signal-plus-noise levels $\lambda_l$. Instead, the Max-Rate solution depends also on the signal-to-noise ratio, which is given by $\lambda_l - 1$. For instance, from (16), if $\lambda_l < 1$, the Max-Rate solution assigns a zero gain to stream $l$, while this is not the case for the MMSE solution (22).

## IV. ROBUST OPTIMAL COMPRESSION

As seen in the previous section, the optimal compression resulting from the solution of problem (9) at the $i$th BS depends on the covariance matrix $\Sigma_{\mathbf{x} | \hat{\mathbf{y}}_S}$ in (13) of the vector of transmitted signals conditioned on the compressed version $\hat{\mathbf{y}}_S$ of the signals received by the BSs in set $\mathcal{S}$. In general, when solving (9), especially in the case in which the optimization is done at the $i$th BS (see Remark 3), it might not be realistic to assume that matrix $\Sigma_{\mathbf{x} | \hat{\mathbf{y}}_S}$ is perfectly known, since depends on the channel matrices of all the BSs in the set $\mathcal{S}$ (see (13)). Motivated by this observation, we propose a robust version of the optimization problem (9) by assuming that, when solving problem (9), only an estimate of $\hat{\Sigma}_{\mathbf{x} | \hat{\mathbf{y}}_S}$ is available that is related to the actual matrix $\Sigma_{\mathbf{x} | \hat{\mathbf{y}}_S}$ as

$$\Sigma_{\mathbf{x} | \hat{\mathbf{y}}_S} = \hat{\Sigma}_{\mathbf{x} | \hat{\mathbf{y}}_S} + \Delta_{\mathbf{x} | \hat{\mathbf{y}}_S}, \tag{23}$$

where $\mathbf{\Delta}_{\mathbf{x}|\hat{\mathbf{y}}_\mathcal{S}} \in \mathcal{H}^{n_M}$ is a deterministic Hermitian matrix that models the estimation error. We assume that the error matrix $\mathbf{\Delta}_{\mathbf{x}|\hat{\mathbf{y}}_\mathcal{S}}$ is only known to belong to a set $\mathcal{U}_\mathbf{\Delta} \subseteq \mathcal{H}^{n_M}$, which models the uncertainty at the $i$th BS regarding matrix $\mathbf{\Sigma}_{\mathbf{x}|\hat{\mathbf{y}}_\mathcal{S}}$.

In general, in order to define the uncertainty set $\mathcal{U}_\mathbf{\Delta}$, one can impose some bounds on given measures of the eigenvalues and/or eigenvectors of matrix $\mathbf{\Delta}_{\mathbf{x}|\hat{\mathbf{y}}_\mathcal{S}}$. Based on the observation that the mutual information $I\left(\mathbf{x}; \hat{\mathbf{y}}_i | \hat{\mathbf{y}}_\mathcal{S}\right)$ in (11) can be written as

$$I\left(\mathbf{x}; \hat{\mathbf{y}}_i | \hat{\mathbf{y}}_\mathcal{S}\right) = f_i\left(\mathbf{\Omega}_i, \tilde{\mathbf{\Delta}}_{\mathbf{x}|\hat{\mathbf{y}}_\mathcal{S}}\right) - \log\det\left(\mathbf{I} + \mathbf{\Omega}_i\right), \tag{24}$$

where we have defined for this section $f_i\left(\mathbf{\Omega}_i, \tilde{\mathbf{\Delta}}_{\mathbf{x}|\hat{\mathbf{y}}_\mathcal{S}}\right) = \log\det\left(\mathbf{I} + \mathbf{\Omega}_i\left(\mathbf{H}_i\hat{\mathbf{\Sigma}}_{\mathbf{x}|\hat{\mathbf{y}}_\mathcal{S}}\mathbf{H}_i^\dagger + \tilde{\mathbf{\Delta}}_{\mathbf{x}|\hat{\mathbf{y}}_\mathcal{S}} + \mathbf{I}\right)\right)$ and $\tilde{\mathbf{\Delta}}_{\mathbf{x}|\mathbf{y}_\mathcal{S}} = \mathbf{H}_i\mathbf{\Delta}_{\mathbf{x}|\hat{\mathbf{y}}_\mathcal{S}}\mathbf{H}_i^\dagger$, we take the approach of bounding the uncertainty on the eigenvalues of $\tilde{\mathbf{\Delta}}_{\mathbf{x}|\hat{\mathbf{y}}_\mathcal{S}}$. This is equivalent to bounding, within some constant, any norm of matrix $\tilde{\mathbf{\Delta}}_{\mathbf{x}|\hat{\mathbf{y}}_\mathcal{S}}$ (see, e.g., [21, Appendix A]). Specifically, we define the uncertainty set $\mathcal{U}_\mathbf{\Delta}$ as the set of Hermitian matrices $\tilde{\mathbf{\Delta}}_{\mathbf{x}|\hat{\mathbf{y}}_\mathcal{S}}$ such that conditions

$$\lambda_{\min}\left(\tilde{\mathbf{\Delta}}_{\mathbf{x}|\hat{\mathbf{y}}_\mathcal{S}}\right) \geq \lambda_{\mathrm{LB}} \ \text{ and } \ \lambda_{\max}\left(\tilde{\mathbf{\Delta}}_{\mathbf{x}|\hat{\mathbf{y}}_\mathcal{S}}\right) \leq \lambda_{\mathrm{UB}} \tag{25}$$

hold for given lower and upper bounds[6] $(\lambda_{\mathrm{LB}}, \lambda_{\mathrm{UB}})$ on the eigenvalues of matrix $\tilde{\mathbf{\Delta}}_{\mathbf{x}|\hat{\mathbf{y}}_\mathcal{S}}$.

Under this model, the problem of deriving the optimal robust compression strategy can be formulated as

$$\underset{\mathbf{\Omega}_i \succeq \mathbf{0}}{\text{maximize}} \ \underset{\tilde{\mathbf{\Delta}}_{\mathbf{x}|\hat{\mathbf{y}}_\mathcal{S}} \in \mathcal{H}^{n_M}}{\min} f_i\left(\mathbf{\Omega}_i, \tilde{\mathbf{\Delta}}_{\mathbf{x}|\hat{\mathbf{y}}_\mathcal{S}}\right) - \log\det\left(\mathbf{I} + \mathbf{\Omega}_i\right)$$

$$\text{s.t.} \begin{cases} f_i\left(\mathbf{\Omega}_i, \tilde{\mathbf{\Delta}}_{\mathbf{x}|\hat{\mathbf{y}}_\mathcal{S}}\right) \leq C_i \\ \lambda_{\min}\left(\tilde{\mathbf{\Delta}}_{\mathbf{x}|\hat{\mathbf{y}}_\mathcal{S}}\right) \geq \lambda_{\mathrm{LB}} \\ \lambda_{\max}\left(\tilde{\mathbf{\Delta}}_{\mathbf{x}|\hat{\mathbf{y}}_\mathcal{S}}\right) \leq \lambda_{\mathrm{UB}} \end{cases} . \tag{26}$$

Problem (26) is not convex and a closed-form solution appears prohibitive. The next theorem derives a solution to the KKT conditions for problem (26), also referred to as a stationary point. It is shown in Appendix A that the KKT conditions are necessary for the optimality of problem (26).

---

[6]Note that when using the additive model (23) a non trivial lower bound $\lambda_{\mathrm{LB}}$ must satisfy $\lambda_{\mathrm{LB}} \geq \lambda_{\min}\left(\mathbf{H}_i\hat{\mathbf{\Sigma}}_{\mathbf{x}|\hat{\mathbf{y}}_\mathcal{S}}\mathbf{H}_i^\dagger\right)$ in order to guarantee that matrix $\mathbf{H}_i\mathbf{\Sigma}_{\mathbf{x}|\hat{\mathbf{y}}_\mathcal{S}}\mathbf{H}_i^\dagger$ is positive semidefinite.

**Theorem 1.** *A stationary point for problem (26) can be found as (10) with matrix $\mathbf{A}_i$ such that $\mathbf{\Omega}_i = \mathbf{A}_i^\dagger \mathbf{A}_i$ is given as (15), where matrix $\mathbf{U}$ is obtained from the eigenvalue decomposition $\mathbf{H}_i \hat{\mathbf{\Sigma}}_{\mathbf{x}|\hat{\mathcal{Y}}_S} \mathbf{H}_i^\dagger + \mathbf{I} = \mathbf{U} \operatorname{diag}\left(\lambda_1, \ldots, \lambda_{n_{B,i}}\right) \mathbf{U}^\dagger$ and the diagonal elements $\alpha_1, \ldots, \alpha_{n_{B,i}}$ are calculated by solving the following mixed integer-continuous problem:*

$$\max_{\mu, \alpha_1 \ldots, \alpha_{n_{B,i}}} g_i^L\left(\alpha_1, \ldots, \alpha_{n_{B,i}}\right) - \sum_{l=1}^{n_{B,i}} \log \det(1 + \alpha_l) \tag{27}$$

$$\text{s.t. } 0 < \mu < 1, \tag{28a}$$

$$\alpha_l \in \mathcal{P}_l(\mu), \ l = 1, \ldots, n_{B,i}, \tag{28b}$$

$$g_i^U\left(\alpha_1, \ldots, \alpha_{n_{B,i}}\right) = C_i, \tag{28c}$$

*where the functions $g_i^L$ and $g_i^U$ are defined as*

$$g_i^L\left(\alpha_1, \ldots, \alpha_{n_{B,i}}\right) = \sum_{l=1}^{n_{B,i}} \log \det(1 + \alpha_l c_l^L), \tag{29}$$

$$\text{and } g_i^U\left(\alpha_1, \ldots, \alpha_{n_{B,i}}\right) = \sum_{l=1}^{n_{B,i}} \log \det(1 + \alpha_l c_l^U), \tag{30}$$

*with $c_l^L = \lambda_l + \lambda_{\text{LB}}$ and $c_l^U = \lambda_l + \lambda_{\text{UB}}$; and the discrete set $\mathcal{P}_l(\mu)$ is defined as*

$$\mathcal{P}_l(\mu) = \begin{cases} \{0\}, & \text{if } Q_l \geq 0, \ S_l \geq 0 \\ \left\{\frac{-Q_l + \sqrt{Q_l^2 - 4S_l}}{2}\right\}, & \text{if } Q_l \geq 0, \ S_l < 0 \\ \left\{\frac{-Q_l \pm \sqrt{Q_l^2 - 4S_l}}{2}, 0\right\}, & \text{if } Q_l < 0, \ S_l \geq 0, \ -\frac{Q_l^2}{4} + S_l \leq 0, \\ \{0\}, & \text{if } Q_l < 0, \ S_l \geq 0, \ -\frac{Q_l^2}{4} + S_l > 0 \\ \left\{\frac{-Q_l + \sqrt{Q_l^2 - 4S_l}}{2}\right\}, & \text{if } Q_l < 0, \ S_l < 0 \end{cases} \tag{31}$$

*with $Q_l$ and $S_l$ given as*

$$Q_l = \frac{c_l^U\left(1 + \mu + (\mu - 1)c_l^L\right)}{\mu c_l^U c_l^L} \ \text{ and } \ S_l = \frac{\mu c_l^U + 1 - c_l^L}{\mu c_l^U c_l^L}. \tag{32}$$

 *Proof:* See Appendix A. ∎

*Remark* 8. The optimal $\alpha_1, \ldots, \alpha_{n_{B,i}}$ for problem (27) can be found from an exhaustive scalar search over $\mu$ between 0 and 1. For each $\mu$, the search for values $\alpha_l$ in problem (27) is restricted to the set $\mathcal{P}_l(\mu)$ that contains at most three elements for $l = 1, \ldots, n_{B,i}$.

In fact, the following corollary shows that, in some special case, the search over parameters $\alpha_l$ is not necessary, since the sets $\mathcal{P}_l(\mu)$ only contain one element.

**Corollary 1.** *If* $\lambda_{\mathrm{UB}} - \lambda_{\mathrm{LB}} < 1$, *a stationary point for problem (26) is given by* $\alpha_l = \frac{\left[-Q_l + \sqrt{Q_l^2 - 4S_l}\right]^+}{2}$ *for* $l = 1, \ldots, n_{B,i}$ *with* $\mu$ *such that the constraint (28c) is satisfied.*

*Proof:* If $\lambda_{\mathrm{UB}} - \lambda_{\mathrm{LB}} < 1$, $\alpha_l$ is computed from (31) as

$$\alpha_l = \begin{cases} \frac{-Q_l + \sqrt{Q_l^2 - 4S_l}}{2}, & \text{if } \mu < \frac{c_l^L - 1}{c_l^U} \\ 0, & \text{if } \mu \geq \frac{c_l^L - 1}{c_l^U} \end{cases}, \tag{33}$$

which entails the claimed result by direct calculation. ∎

*Remark 9.* If $\lambda_{\mathrm{UB}} = \lambda_{\mathrm{LB}} = 0$, the solution to problem (27) is unique and reduces to the solution (16) obtained if matrix $\Sigma_{\mathbf{x}|\hat{\mathbf{y}}_S}$ is perfectly known at BS $i$. This shows that the proposed robust solution reduces to the optimal design for the case in which $\Sigma_{\mathbf{x}|\hat{\mathbf{y}}_S}$ is perfectly known to BS $i$.

*Proof:* It follows by substituting $c_l^U = c_l^L = \lambda_l$ into (33). ∎

*Remark 10.* As shown in Appendix A, any values of $\mu, \alpha_1, \ldots, \alpha_{n_{B,i}}$ that satisfy the constraints (28a)-(28c) are a solution to the KKT conditions for the robust problem (26).

## V. Joint BS Selection and Compression via Sparsity-Inducing Optimization

In order to operate the network efficiently, it is generally advantageous to let only a subset $\mathcal{S} \subseteq \mathcal{N}_B$ of the $N_B$ available BSs communicate to the cloud decoder in a given time-slot. This is the case in scenarios in which different BSs share the same backhaul resources [27] or energy consumption and green networking are critical issues [24]. Therefore, under this assumption, the system design entails the choice of the subset $\mathcal{S}$, along with that of the compression test channels $p(\hat{\mathbf{y}}_i|\mathbf{y}_i)$ for $i \in \mathcal{S}$. In general, this BS selection requires an exhaustive search of exponential complexity in the number $N_B$ of BSs. Here, inspired by [23], we propose an efficient approach based on the addition of a sparsity-inducing term to the objective function.

To elaborate, we associate a cost $q_i$ per spectral unit resource (i.e., per discrete-time channel use) to the $i$th BS. This measures the relative cost per spectral resource of activating the $i$th BS over the revenue per bit. To fix the ideas, consider a single cell with one MBS and $N_B - 1$ HBSs. We assume that the HBSs share the same total backhaul capacity $C_H$ to the cloud decoder, as is the case if the HBSs communicate to the cloud decoder via a shared wireless link [27].

Assuming that the MBS is active, we are interested in selecting a subset of the HBSs so as to provide additional information to the cloud decoder under the given total backhaul constraint. Note that the solution proposed here can also be generalized to more complex systems with multiple cells.

Let $\mathcal{S}_{\mathcal{M}} = \{1\}$ and $\mathcal{S}_{\mathcal{H}} = \{2, \ldots, N_B\}$ denote the set that includes the MBS and the HBSs, respectively. Assuming that the Gaussian test channel (10) is employed at each $i$th BS with a given covariance matrix $\Omega_i \succeq \mathbf{0}$, the joint problem of HBS selection and compression via sparsity-inducing optimization is formulated as

$$\underset{\{\Omega_i \succeq \mathbf{0}\}_{i \in \mathcal{S}_{\mathcal{H}}}}{\text{maximize}} \; I\left(\mathbf{x}; \hat{\mathbf{y}}_{\mathcal{S}_{\mathcal{H}}} | \hat{\mathbf{y}}_1\right) - q_H \sum_{i \in \mathcal{S}_{\mathcal{H}}} \mathbf{1}\left(\|\Omega_i\|_F > 0\right) \tag{34}$$

$$\text{s.t. } I\left(\mathbf{y}_{\mathcal{S}_{\mathcal{H}}}; \hat{\mathbf{y}}_{\mathcal{S}_{\mathcal{H}}} | \hat{\mathbf{y}}_1\right) \leq C_H,$$

where $\mathbf{1}(\cdot)$ is the indicator function, which takes 1 if the argument state is true and 0 otherwise, and we have assumed that $q_2 = \ldots = q_{N_B} = q_H$ for simplicity. In (34), we have conditioned on $\hat{\mathbf{y}}_1$ to account for the fact that the MBS is assumed to be active. Note that the second term in the objective of problem (34) is the $\ell_0$-norm of vector $[\, \text{tr}(\Omega_2) \; \cdots \; \text{tr}(\Omega_{N_B}) \,]$. If the cost $q_H$ is large enough, this term forces the solution to set some of the matrices $\Omega_i$ to zero, thus keeping the corresponding $i$th HBS inactive. As is standard practice, in order to avoid the non-smoothness induced by the $\ell_0$-norm, we modify problem (34) by replacing the $\ell_0$-norm with the $\ell_1$-norm of the same vector. We thus reformulate problem (34) as follows:

$$\underset{\{\Omega_i \succeq \mathbf{0}\}_{i \in \mathcal{S}_{\mathcal{H}}}}{\text{maximize}} \; f\left(\Omega_2, \ldots, \Omega_{N_B}\right) \tag{35}$$

$$\text{s.t. } g\left(\Omega_2, \ldots, \Omega_{N_B}\right) \leq C_H,$$

where we have defined for this section $f\left(\Omega_2, \ldots, \Omega_{N_B}\right) = I\left(\mathbf{x}; \hat{\mathbf{y}}_{\mathcal{S}_{\mathcal{H}}} | \hat{\mathbf{y}}_1\right) - q_H \sum_{i \in \mathcal{S}_{\mathcal{H}}} \text{tr}\left(\Omega_i\right)$ and $g\left(\Omega_2, \ldots, \Omega_{N_B}\right) = I\left(\mathbf{y}_{\mathcal{S}_{\mathcal{H}}}; \hat{\mathbf{y}}_{\mathcal{S}_{\mathcal{H}}} | \hat{\mathbf{y}}_1\right)$. An explicit expansion for $f\left(\Omega_2, \ldots, \Omega_{N_B}\right)$ and $g\left(\Omega_2, \ldots, \Omega_{N_B}\right)$ as a function of $\Omega_2, \ldots, \Omega_{N_B}$ can be easily obtained and is not regarded now.

Based on this formulation, we propose a two-phase approach to the problem of joint HBS selection and compression in Algorithm 2. As illustrated in the table, in the first phase we execute the block-coordinate ascent algorithm [22] to tackle problem (35). As a result, we obtain a subset $\mathcal{S}_{\mathcal{H}}^* \subseteq \mathcal{S}_{\mathcal{H}}$ of HBSs with nonzero $\Omega_i$. In the second phase, the block-coordinate ascent algorithm

---

**Algorithm 2** Two-Phase Joint HBS Selection and Compression Algorithm

---

Phase 1. Solve problem (35) via the block-coordinate ascent algorithm:

i) Initialize $n = 0$ and $\boldsymbol{\Omega}_2^{(n)} = \ldots = \boldsymbol{\Omega}_{N_B}^{(n)} = \mathbf{0}$;

ii) For $i = 2, \ldots, N_B$, update $\boldsymbol{\Omega}_i^{(n)}$ as a solution of the following problem.

$$\underset{\boldsymbol{\Omega}_i \succeq \mathbf{0}}{\text{maximize}} \; f\left(\boldsymbol{\Omega}_i, \boldsymbol{\Omega}_{\{2,\ldots,i-1\}}^{(n-1)}, \boldsymbol{\Omega}_{\{i+1,\ldots,N_B\}}^{(n)}\right) \tag{36}$$

$$\text{s.t.} \; g\left(\boldsymbol{\Omega}_i, \boldsymbol{\Omega}_{\{2,\ldots,i-1\}}^{(n-1)}, \boldsymbol{\Omega}_{\{i+1,\ldots,N_B\}}^{(n)}\right) \leq C_H.$$

iii) Repeat step ii) if some convergence criterion is not satisfied and stop otherwise. Once the algorithm has terminated, denote the obtained $\boldsymbol{\Omega}_i$ by $\boldsymbol{\Omega}_i^*$ for $i = 2, \ldots, N_B$, and set $\mathcal{S}_{\mathcal{H}}^* = \{i \in \mathcal{S}_{\mathcal{H}} : \boldsymbol{\Omega}_i^* \neq \mathbf{0}\}$.

Phase 2. Apply the block-coordinate ascent algorithm to problem (35) with $q_H = 0$ and considering only the HBSs in set $\mathcal{S}_{\mathcal{H}}^*$.

---

is run only on the subset $\mathcal{S}_{\mathcal{H}}^*$ by setting $\boldsymbol{\Omega}_i = \mathbf{0}$ for all $i \notin \mathcal{S}_{\mathcal{H}}^*$ and $q_H = 0$. It is noted that with $q_H = 0$, the block-coordinate ascent method used here is the same as proposed in [11, Sec. IV].

It remains to discuss how to solve problem (36) at step ii) of the proposed algorithm. Note that this corresponds to the update of $\boldsymbol{\Omega}_i$ when all the other variables $\boldsymbol{\Omega}_{\mathcal{S}_{\mathcal{H}} \setminus \{i\}}$ are fixed to the values obtained from the previous iterations. The global maximum of problem (36) can be obtained as shown in Theorem 2.

**Theorem 2.** *A solution to problem (36) is given by (10), with matrix $\mathbf{A}_i$ such that $\boldsymbol{\Omega}_i = \mathbf{A}_i^\dagger \mathbf{A}_i$ is given as (15), where we have the eigenvalue decomposition $\boldsymbol{\Sigma}_{\mathbf{y}_i | \hat{\mathbf{y}}_{\mathcal{N}_{\mathcal{B}} \setminus \{i\}}} = \mathbf{U} \text{diag}(\lambda_1, \ldots, \lambda_{n_{B,i}}) \mathbf{U}^\dagger$, and matrix $\boldsymbol{\Sigma}_{\mathbf{y}_i | \hat{\mathbf{y}}_{\mathcal{N}_{\mathcal{B}} \setminus \{i\}}}$ is given as*

$$\boldsymbol{\Sigma}_{\mathbf{y}_i | \hat{\mathbf{y}}_{\mathcal{N}_{\mathcal{B}} \setminus \{i\}}} = \mathbf{I} + \mathbf{H}_i \mathbf{R}_i^{-1} \boldsymbol{\Sigma}_{\mathbf{x} | \hat{\mathbf{y}}_1} \mathbf{H}_i^\dagger, \tag{37}$$

*with $\mathbf{R}_i = \mathbf{I} + \boldsymbol{\Sigma}_{\mathbf{x} | \hat{\mathbf{y}}_1} \sum_{j \in \mathcal{S}_{\mathcal{H}} \setminus \{i\}} \mathbf{H}_j^\dagger \left(\mathbf{I} + \boldsymbol{\Omega}_j\right)^{-1} \boldsymbol{\Omega}_j \mathbf{H}_j$. The diagonal elements $\alpha_1, \ldots, \alpha_{n_{B,i}}$ are obtained as $\alpha_l = \alpha_l(\mu^*)$, with*

$$\alpha_l(\mu) = \frac{\left[ -\lambda_l \mu - q_H'(\lambda_l + 1) + \sqrt{(\lambda_l \mu + q_H'(\lambda_l + 1))^2 - 4 q_H' \lambda_l \left((\mu - 1)\lambda_l + q_H' + 1\right)} \right]^+}{2 q_H' \lambda_l}, \tag{38}$$

*for $l = 1, \ldots, n_{B,i}$ with $q_H' = \log_e 2 \cdot q_H$. The Lagrange multiplier $\mu^*$ is obtained as follows: if*

$h_i(0) \leq \bar{C}_i$, *where* $\bar{C}_i$ *is given by*

$$\bar{C}_i = C_i - \log \det \mathbf{R}_i - \sum_{j \in \mathcal{S}_{\mathcal{H} \setminus \{i\}}} \log \det (\mathbf{I} + \mathbf{\Omega}_j), \tag{39}$$

*then* $\mu^* = 0$; *otherwise,* $\mu^*$ *is unique value* $\mu \geq 0$ *such that* $h_i(\mu) = \bar{C}_i$ *where* $h_i(\mu) = \sum_{l=1}^{n_{B,i}} \log (1 + \lambda_l \alpha_l(\mu))$.

    *Proof:* The proof is given in Appendix B. ∎

*Remark* 11. If $q_H = 0$, the solution (38) reduces to (16) as derived in [11].

## VI. NUMERICAL RESULTS

In this section, we present numerical results in order to validate and complement the analysis. The sum-rate performance is evaluated for single-cell and multi-cell systems with perfect and imperfect side information. It is assumed that each cell with cell-radius $R_{\text{cell}}$ has a single MBS and multiple HBSs. The MBS is located at cell center while the HBS and MS are randomly dropped within a circular cell according to uniform distribution. All channel elements of $\mathbf{H}_{i,j}$ are i.i.d. circularly symmetric complex Gaussian variables with zero mean and variance $(D_0/d_{i,j})^\nu$, where the path-loss exponent $\nu$ is chosen as 3.5 and $d_{i,j}$ is the distance from MS $j$ to BS $i$. The reference distance $D_0$ is set to half of cell-radius, i.e., $D_0 = R_{\text{cell}}/2$. For simplicity, each MS uses single antenna, i.e., $n_{M,i} = 1$ with transmit power $P_{\text{tx}}$ such that the aggregated transmit vector $\mathbf{x}$ has a covariance of $\mathbf{\Sigma}_{\mathbf{x}} = P_{\text{tx}} \mathbf{I}$. Then, the SNR is defined as $P_{\text{tx}}$ since we have assumed unit variance noise in Sec. II.

For Sec. VI-A and VI-B, we assume the greedy ordering described in Sec. III-A. Moreover, we assume that the MBS is connected to the cloud decoder via a backhaul link of capacity $C$ while the HBSs' backhaul is of capacity equal to a fraction of $C$, namely $\omega C$ with $0 < \omega \leq 1$. The setup for Sec. VI-C will be that of Sec. V (to be described below).

### A. Compressions with Perfect Side Information

In this subsection, we compare the compression schemes discussed in Sec. III with perfect side information, i.e., under the assumption that each BS knows the relevant covariance matrix $\mathbf{\Sigma}_{\mathbf{x}|\hat{\mathbf{y}}_\mathcal{S}}$. We consider three hexagonal cells, in which each cell has 1 MBS, 2 HBSs and 3 MSs such that $N_B = 9$ and $N_M = 9$. In Fig. 3, we plot the per-MS sum-rate performance[7], averaged over network topology and fading, as a function of the fractional backhaul capacity $\omega$ of the HBSs. Each BS has $n_{B,i} = 8$ antennas and the average received SNR is set to $P_{\text{tx}} = -5\text{dB}$ at the reference distance of $D_0$. The MBS backhaul link is of capacity $C = 15$ bps/Hz. It is seen that Max-Rate compression outperforms the MMSE-based schemes. Moreover, leveraging side information at the cloud decoder results in significant gains, especially for Max-Rate compression. It is also interesting to observe that indirect MMSE methods can be outperformed by direct MMSE methods. We remark that indirect MMSE methods always have better MMSE performance, but, as shown here, they might lead to a performance loss in terms of achievable rate. Our results show that the performance comparison between indirect and direct MMSE methods depends on the specific network, and that, as expected, indirect method can also be preferable in some scenarios (not shown here). Moreover, for further comparison, Fig. 3 presents the rate performance of a standard decode-and-forward scheme in which each MS is assigned to the BS to which the received power is the maximum. It is noted that the decode-and-forward strategy shows poor performance especially for small ratio $\omega$ since the performance of the MSs connected to the HBSs is limited, not only by the interference caused by the MSs assigned to other BSs but also by the HBS backhaul capacity.

We now further elaborate on the impact of side information on the performance of Max-Rate compression. To this end, in Fig. 4 we plot the average per-MS rate versus the SNR $P_{\text{tx}}$ for a network with $C = 10$ bps/Hz and $\omega$=0.5. It is seen that leveraging the availability of side information becomes especially significant as the number of BS antennas becomes large and/or as the SNR grows. This is since an increased number of receive antennas implies that the received signal lies in a larger dimensional space, which calls for more effective compression strategies. Moreover, the compression noise affects the performance more as the noise level on the received signal decreases.

### B. Imperfect Side Information

In this subsection, we present numerical results for the case in which there is uncertainty on the conditional covariance matrix $\Sigma_{\mathbf{x}|\hat{\mathbf{y}}_S}$. According to the considered uncertainty model, matrix

---

[7]The per-MS rate is calculated by normalizing the sum-rate by the number $N_M$ of MSs

$\hat{\Sigma}_{\mathbf{x}|\hat{\mathbf{y}}_S}$ is assumed to be known at the $i$th BS with an uncertainty matrix $\tilde{\mathbf{\Delta}}_{\mathbf{x}|\hat{\mathbf{y}}_S} = \mathbf{H}_i \mathbf{\Delta}_{\mathbf{x}|\hat{\mathbf{y}}_S} \mathbf{H}_i^\dagger$ (see Sec. IV) whose eigenvalues are limited in the range (25) for given bounds $(\lambda_{\mathrm{LB}}, \lambda_{\mathrm{UB}})$. On the simulations, we have generated the eigenvectors of $\tilde{\mathbf{\Delta}}_{\mathbf{x}|\hat{\mathbf{y}}_S}$ randomly according to isotropic distribution on the column space of $\mathbf{H}_i$ and the eigenvalues uniformly in the set (25) where $\lambda_{\mathrm{UB}} = \lambda_{\min}\left(\mathbf{H}_i \mathbf{\Sigma}_{\mathbf{x}|\hat{\mathbf{y}}_S} \mathbf{H}_i^\dagger\right)$ and $\lambda_{\mathrm{LB}} = -\lambda_{\min}\left(\mathbf{H}_i \mathbf{\Sigma}_{\mathbf{x}|\hat{\mathbf{y}}_S} \mathbf{H}_i^\dagger\right)$, respectively. This implies that the uncertainty on the eigenvalues is the maximum (symmetric) uncertainty consistent with the positive semi-definiteness of matrix $\mathbf{H}_i \hat{\mathbf{\Sigma}}_{\mathbf{x}|\hat{\mathbf{y}}_S} \mathbf{H}_i^\dagger$ (See Sec. IV).

In Fig. 5, per-MS sum-rate performance is evaluated for a single-cell with $N_B = 4$ (one MBS and three randomly placed HBSs), $N_M = 8$, $n_{B,i} = 2$, $\omega = 0.5$ and $P_{\mathrm{tx}} = 10\mathrm{dB}$ versus the MBS backhaul capacity $C$. For reference, we plot the performance attained by a variation of the Max-Rate scheme that ignores side information[8] and that of a scheme that operates by assuming that $\hat{\mathbf{\Sigma}}_{\mathbf{x}|\hat{\mathbf{y}}_S}$ is the true covariance matrix. Note that, in this case, the backhaul constraint (9) can be violated, which implies that the cloud decoder cannot recover the corresponding compressed signal $\hat{\mathbf{y}}_i$ (labeled as "imperfect SI" in the figure). The figure shows that assuming the incorrect matrix $\hat{\mathbf{\Sigma}}_{\mathbf{x}|\hat{\mathbf{y}}_S}$ as being true can result in a severe performance degradation. However, this performance loss can be overcome by adopting the proposed robust algorithm, which shows intermediate performance between the ideal setting with perfect side information and that with no side information. We also observe the more pronounced performance gain of the proposed robust solution for a larger backhaul link capacity. In a similar vein, in Fig. 6, we investigate the effect of the number $N_B - 1$ of HBSs. It is seen that, as the number of BSs grows, leveraging side information provides more relevant gains, so that, even assuming imperfect side information can be useful. As in Fig. 6, the proposed algorithm shows a rate performance close to the ideal case of perfect side information.

## C. Joint HBS Selection and Compression

In this subsection, we consider the single-cell setup of Sec. V, where we aim at selecting a subset of the $N_B - 1$ HBSs under total backhaul constraint $C_H$. We compare the proposed two-phase approach (Algorithm 2) with:

i) $N_H^*$ *Exhaustive Search*: the scheme selects $N_H^*$ HBSs that maximize the sum-rate via

---

[8]This is easily obtained from (15) and (16) by substituting the covariance matrix $\mathbf{\Sigma}_{\mathbf{x}|\hat{\mathbf{y}}_S}$ with $\mathbf{\Sigma}_{\mathbf{x}}$.

exhaustive search, where $N_H^*$ is the cardinality of the set $\mathcal{S}_{\mathcal{H}}^*$ obtained after the first phase of the two-phase algorithm;

ii) $N_H^*$ *Local*: the scheme selects the $N_H^*$ HBSs with the largest value $C_i^{\text{local}}$, where $C_i^{\text{local}}$ is the capacity from the $N_M$ MSs to BS $i$, i.e., $C_i^{\text{local}} = \log \det \left( \mathbf{I} + \mathbf{H}_i \mathbf{\Sigma_x} \mathbf{H}_i^\dagger \right)$ [28]. Note that this criterion is local in that it does not account for the correlation between the signals received by different HBSs.

iii) $N_H^*$ *Random*: the scheme randomly selects $N_H^*$ HBSs.

We consider a practical scenario in which the HBSs and MSs are divided into two groups: $N_B^1$ HBSs and $N_M^1$ MSs in group 1 and $N_B^2$ HBSs and $N_M^2$ MSs in group 2 such that $N_B^1 + N_B^2 = N_B$ and $N_M^1 + N_M^2 = N_M$. The HBSs and MSs in group 1 are uniformly distributed within the whole cell, while those in group 2 are distributed in smaller cell overlaid on the macrocell at hand and with radius $R_{\text{spot}} < R_{\text{cell}}$, which models a *"hot spot"* such as a building or a public space. Fig. 7 presents the per-MS sum-rate versus the ratio $R_{\text{spot}}/R_{\text{cell}}$ with $N_B = 13$ ($N_B^1 = 6$, $N_B^2 = 6$), $N_M = 14$ ($N_M^1 = 8$, $N_M^2 = 6$), $n_{B,i} = 8$, $C = 20$ bps/Hz, $C_H = 300$ bps/Hz and $q_H = 100$. From the figure, it is observed that the $N_H^*$ local approach provides a performance close to the $N_H^*$ exhaustive search approach when the size of the *hot spot* is large. In fact, in this case, the correlation between pairs of signals received by different HBSs tends to be similar given the symmetry of the network topology. However, for sufficiently small *hot spot* size, the performance loss of the $N_H^*$ local approach becomes significant, while the proposed two-phase scheme still shows a performance almost identical to that of the $N_H^*$ exhaustive search scheme (which requires a search over $\binom{12}{6} = 924$ combinations[9] of HBSs). This is because, in this case, signals received by HBSs in the smaller *hot spot* tend to be more correlated, and thus it is more advantageous to select the HBSs judiciously in order to increase the sum-rate.

## VII. Conclusions

In this work, we have studied distributed compression schemes for the uplink of cloud radio access networks. We proposed a robust compression scheme for a practical scenario with inaccurate statistical information about the correlation among the BSs' signals. The scheme is based on a deterministic worst-case problem formulation and the solution of the corresponding

---

[9]In the simulation, it was observed that the average number $N_H^*$ of activated HBSs is about 6 for the simulated configurations.

KKT conditions. Via numerical results, we have demonstrated that, while errors in the statistical model of the side information make distributed source coding strategy virtually useless, the proposed robust compression scheme allows to tolerate sizable errors without drastic performance degradation and while still reaping the benefits of distributed source coding. In this regard, we remark that the robust strategy could be further improved in at least two ways, which are subject of current work. First, one could deploy a layered compression strategy that attempts to opportunistically leverage a more advantageous side information (see, e.g., [31]). Second, one could enhance the decoding operation by performing joint decompression and decoding as discussed in [5] for a simplified model. Moreover, we have addressed the issue of selecting a subset of BSs with the aim of improving the energy efficiency of the network. This was done by proposing a joint BS selection and compression approach, in which a sparsity-inducing term is introduced into the objective function. It was verified that the proposed joint BS scheduling and compression method shows performance close to exhaustive search.

## Appendix A

## Proof of Theorem 1

Since the problem (26) involves infinitely many inequality constraints, we first convert it into a problem with finite number of inequalities following the standard robust optimization approach reviewed in [29]. This step will also allow us to eliminate variable $\tilde{\boldsymbol{\Delta}}_{\mathbf{x}|\hat{\mathbf{y}}_S}$ from the problem formulation. We then show that the KKT conditions are necessary for optimality. Finally, we formulate the KKT conditions and verify that a solution to problem (27) also satisfies the KKT conditions.

**Lemma 3.** *Problem (26) is equivalent to the problem*

$$\underset{\boldsymbol{\Omega}_i \succeq \mathbf{0}}{\text{maximize}} \; f_i\left(\boldsymbol{\Omega}_i, \lambda_{\mathrm{LB}}\mathbf{I}\right) - \log\det\left(\boldsymbol{\Omega}_i + \mathbf{I}\right) \tag{40}$$

$$\text{s.t.} \; f_i\left(\boldsymbol{\Omega}_i, \lambda_{\mathrm{UB}}\mathbf{I}\right) - C_i \leq 0.$$

*Proof:* First, we consider the epigraph form of problem (26) (see, e.g., [21, Sec. 4.1.3]):

$$\underset{\boldsymbol{\Omega}_i \succeq \mathbf{0}, \, t, \, \tilde{\boldsymbol{\Delta}}_{\mathbf{x}|\hat{\mathbf{y}}_S} \in \mathcal{H}^{n_M} \text{ s.t. } (25)}{\text{maximize}} \; t - \log\det\left(\mathbf{I} + \boldsymbol{\Omega}_i\right) \tag{41}$$

$$\text{s.t.} \; \max\left\{t - f_i\left(\boldsymbol{\Omega}_i, \tilde{\boldsymbol{\Delta}}_{\mathbf{x}|\hat{\mathbf{y}}_S}\right), \; f_i\left(\boldsymbol{\Omega}_i, \tilde{\boldsymbol{\Delta}}_{\mathbf{x}|\hat{\mathbf{y}}_S}\right) - C_i\right\} \leq 0.$$

Then, we observe that problem (41) is equivalent to the following problem with one inequality constraint:

$$\underset{\boldsymbol{\Omega}_i \succeq \mathbf{0},\, t}{\text{maximize}} \; t - \log\det\left(\mathbf{I} + \boldsymbol{\Omega}_i\right) \tag{42}$$

$$\text{s.t.} \underset{\tilde{\boldsymbol{\Delta}}_{\mathbf{x}|\hat{\mathbf{y}}_{\mathcal{S}}} \text{ s.t. (25)}}{\max} \max\left\{ t - f_i\left(\boldsymbol{\Omega}_i, \tilde{\boldsymbol{\Delta}}_{x|\hat{y}_{\mathcal{S}}}\right),\; f_i\left(\boldsymbol{\Omega}_i, \tilde{\boldsymbol{\Delta}}_{x|\hat{y}_{\mathcal{S}}}\right) - C_i \right\} \leq 0.$$

Since the inequality constraint in (42) can be written as

$$\max\left\{ \begin{array}{l} t - \underset{\tilde{\boldsymbol{\Delta}}_{\mathbf{x}|\hat{\mathbf{y}}_{\mathcal{S}}} \text{ s.t. (25)}}{\min} f_i\left(\boldsymbol{\Omega}_i, \tilde{\boldsymbol{\Delta}}_{x|\hat{y}_{\mathcal{S}}}\right), \\ \underset{\tilde{\boldsymbol{\Delta}}_{\mathbf{x}|\hat{\mathbf{y}}_{\mathcal{S}}} \text{ s.t. (25)}}{\max} f_i\left(\boldsymbol{\Omega}_i, \tilde{\boldsymbol{\Delta}}_{x|\hat{y}_{\mathcal{S}}}\right) - C_i \end{array} \right\} \leq 0, \tag{43}$$

in order to proceed, we need to maximize and minimize the function $f_i$ with respect to $\tilde{\boldsymbol{\Delta}}_{\mathbf{x}|\hat{\mathbf{y}}_{\mathcal{S}}}$ for given $\boldsymbol{\Omega}_i$ under constraint (25). To this end, note that function $f_i$ can be written as

$$f_i\left(\boldsymbol{\Omega}_i, \tilde{\boldsymbol{\Delta}}_{x|\hat{y}_{\mathcal{S}}}\right) = f_i\left(\boldsymbol{\Omega}_i, \mathbf{0}\right) + \log\det\left(\mathbf{I} + \mathbf{K}_i \tilde{\boldsymbol{\Delta}}_{\mathbf{x}|\mathbf{y}_{\mathcal{S}}}\right) \tag{44}$$

where $\mathbf{K}_i = \left(\mathbf{I} + \boldsymbol{\Omega}_i \left(\mathbf{H}_i \hat{\boldsymbol{\Sigma}}_{\mathbf{x}|\hat{\mathbf{y}}_{\mathcal{S}}} \mathbf{H}_i^{\dagger} + \mathbf{I}\right)\right)^{-1} \boldsymbol{\Omega}_i$. The matrix $\tilde{\boldsymbol{\Delta}}_{\mathbf{x}|\hat{\mathbf{y}}_{\mathcal{S}}}$ affects only the second term, which can be also expressed as

$$\log\det\left(\mathbf{I} + \mathbf{K}_i \tilde{\boldsymbol{\Delta}}_{\mathbf{x}|\mathbf{y}_{\mathcal{S}}}\right) = \sum_{l=1}^{n_{B,i}} \log\left(1 + \lambda_l\left(\mathbf{K}_i \tilde{\boldsymbol{\Delta}}_{\mathbf{x}|\mathbf{y}_{\mathcal{S}}}\right)\right), \tag{45}$$

where $\lambda_l(\mathbf{X})$ represents the $l$th largest eigenvalue of $\mathbf{X}$. Finally, using the following eigenvalue inequalities [30],

$$\lambda_l\left(\mathbf{K}_i\right) \lambda_{\min}\left(\tilde{\boldsymbol{\Delta}}_{\mathbf{x}|\mathbf{y}_{\mathcal{S}}}\right) \leq \lambda_l\left(\mathbf{K}_i \tilde{\boldsymbol{\Delta}}_{\mathbf{x}|\mathbf{y}_{\mathcal{S}}}\right) \leq \lambda_l\left(\mathbf{K}_i\right) \lambda_{\max}\left(\tilde{\boldsymbol{\Delta}}_{\mathbf{x}|\mathbf{y}_{\mathcal{S}}}\right), \tag{46}$$

for $l = 1, \ldots, n_{B,i}$, the optimal values for the maximization and minimization of $f_i$ are obtained by setting $\tilde{\boldsymbol{\Delta}}_{\mathbf{x}|\hat{\mathbf{y}}_{\mathcal{S}}} = \lambda_{\mathrm{UB}}\mathbf{I}$ and $\tilde{\boldsymbol{\Delta}}_{\mathbf{x}|\hat{\mathbf{y}}_{\mathcal{S}}} = \lambda_{\mathrm{LB}}\mathbf{I}$, respectively. This leads to problem (40). ∎

Problem (40) is not convex due to the non-convexity of constraint set. In the next two lemmas, we list some necessary conditions for the optimality of problem (40).

**Lemma 4.** *The KKT conditions are necessary conditions for optimality in problem (40).*

*Proof:* It follows from [22] and direction calculation. ∎

**Lemma 5.** *At any optimal point $\boldsymbol{\Omega}_i^*$ for problem (40), the backhaul capacity should be fully utilized, i.e., $f_i\left(\boldsymbol{\Omega}_i^*, \lambda_{\mathrm{UB}}\mathbf{I}\right) = C_i$.*

*Proof:* Suppose that $\Omega_i^0$ is optimal but does not fully utilize the backhaul capacity. Then, we can set $\Omega_i = \eta \Omega_i^0$ with some $\eta > 1$ so as to increase the objective function in (40) without violating the backhaul capacity constraint of problem (40). Thus, $\Omega_i^0$ cannot be the optimal solution. ∎

Now, without loss of optimality, we can consider only the points satisfying the necessary conditions described in Lemmas 4 and 5. To this end, we first define the Lagrangian function for the problem (40) as

$$\mathcal{L}\left(\Omega_i, \mu, \Upsilon\right) = f_i\left(\Omega_i, \lambda_{\mathrm{LB}}\mathbf{I}\right) - \log\det\left(\Omega_i + \mathbf{I}\right) - \mu\left(f_i\left(\Omega_i, \lambda_{\mathrm{UB}}\mathbf{I}\right) - C_i\right) + \mathrm{tr}\left(\Upsilon\Omega_i\right), \quad (47)$$

with Lagrange multipliers $\mu \geq 0$ and $\Upsilon \succeq \mathbf{0}$. Then, the KKT conditions with full-backhaul constraint are written as (see, e.g., [21, Sec. 5.5.3])

$$\frac{\partial \mathcal{L}}{\partial \Omega_i} = \mathbf{0}, \quad (48a)$$

$$\mathrm{tr}\left(\Upsilon\Omega_i\right) = 0, \quad (48b)$$

$$\text{and } \; f_i\left(\Omega_i, \lambda_{\mathrm{UB}}\mathbf{I}\right) - C_i = 0 \quad (48c)$$

where the derivative $\partial\mathcal{L}/\partial\Omega_i$ is computed as

$$\frac{\partial \mathcal{L}}{\partial \Omega_i} = \left(\mathbf{H}_i\hat{\boldsymbol{\Sigma}}_{\mathbf{x}|\hat{\mathbf{y}}_S}\mathbf{H}_i^\dagger + (\lambda_{\mathrm{LB}}+1)\,\mathbf{I}\right)\left(\mathbf{I} + \Omega_i\left(\mathbf{H}_i\hat{\boldsymbol{\Sigma}}_{\mathbf{x}|\hat{\mathbf{y}}_S}\mathbf{H}_i^\dagger + (\lambda_{\mathrm{LB}}+1)\,\mathbf{I}\right)\right)^{-1} - (\Omega_i + \mathbf{I})^{-1}$$
$$- \mu\left(\mathbf{H}_i\hat{\boldsymbol{\Sigma}}_{\mathbf{x}|\hat{\mathbf{y}}_S}\mathbf{H}_i^\dagger + (\lambda_{\mathrm{UB}}+1)\,\mathbf{I}\right)\left(\mathbf{I} + \Omega_i\left(\mathbf{H}_i\hat{\boldsymbol{\Sigma}}_{\mathbf{x}|\hat{\mathbf{y}}_S}\mathbf{H}_i^\dagger + (\lambda_{\mathrm{UB}}+1)\,\mathbf{I}\right)\right)^{-1} + \Upsilon. \quad (49)$$

We now set $\Omega_i$ as in (15) and argue that selecting the eigenvalues $\alpha_1, \ldots, \alpha_{n_{B,i}}$ as in (27)-(28c) provides a solution to the KKT conditions (48a)-(48c). Specifically, with the choice (15), the necessary conditions (48a)-(48c) can be written as

$$\frac{c_l^L}{1 + \alpha_l c_l^L} - \frac{1}{1 + \alpha_l} - \frac{\mu c_l^U}{1 + \alpha_l c_l^U} + \theta_l = 0, \; l = 1, \ldots, n_{B,i}, \quad (50a)$$

$$\theta_l\alpha_l = 0, \; \theta_l \geq 0, \; l = 1, \ldots, n_{B,i}, \quad (50b)$$

$$\text{and } \; g_i^U\left(\alpha_1, \ldots, \alpha_{n_{B,i}}\right) - C_i = 0, \quad (50c)$$

with $\theta_l \geq 0$ for $l = 1, \ldots, n_{B,i}$. We note that, similar to Lemmas 4 and 5, the conditions

(50a)-(50c) can be shown to be necessary for the optimality of the following problem.

$$\underset{\alpha_1 \geq 0, \ldots, \alpha_{n_{B,i}} \geq 0}{\text{maximize}} \; g_i^L \left( \alpha_1, \ldots, \alpha_{n_{B,i}} \right) - \sum_{l=1}^{n_{B,i}} \log \det \left( 1 + \alpha_l \right), \tag{51}$$

$$\text{s.t. } g_i^U \left( \alpha_1, \ldots, \alpha_{n_{B,i}} \right) - C_i = 0.$$

But, according to the Weierstrass Theorem [22], the problem (51) has a solution due to the compact constraint set. Thus, we can find parameters $\alpha_i, \ldots, \alpha_{n_{B,i}}$ satisfying the KKT conditions (50a)-(50c) with a proper choice of $\mu$.

The discussion above shows that any solution of problem (51) provides a solution to the KKT conditions (50a)-(50c). Moreover, we show that $\alpha_l$ must lie in the set $\mathcal{P}_l(\mu)$ with $\mu \in (0, 1)$ in order to satisfy the conditions (50a)-(50c), and thus we can limit the domain of the optimization (51) as done in (27)-(28c). This is shown next. Firstly, from the following lemma, the search region for $\mu$ can be restricted to the interval $\mu \in (0, 1)$.

**Lemma 6.** *For $\mu = 0$ and $\mu \geq 1$, the conditions (50a)-(50c) cannot be satisfied simultaneously.*

*Proof:* With $\mu = 0$, it is impossible to satisfy (50a) and (50b) simultaneously. For $\mu \geq 1$, (50a) does not hold with nonnegative $\alpha_l$. ∎

**Lemma 7.** *A value of $\alpha_l$ with $\alpha_l \notin \mathcal{P}_l$ cannot satisfy the conditions (50a)-(50c).*

*Proof:* In order for (50a) and (50b) to hold together, parameter $\alpha_l$ should be such that

$$\alpha_l^2 + Q_l \alpha_l + S_l = 0, \text{ if } \alpha_l > 0, \tag{52}$$

$$\text{and } \alpha_l^2 + Q_l \alpha_l + S_l \geq 0, \text{ if } \alpha_l = 0. \tag{53}$$

By direct calculation, it follows that candidate $\alpha_l \in \mathcal{P}_l(\mu)$ must hold in order to satisfy both (52) and (53). ∎

## Appendix B

### Proof of Theorem 2

The following lemma provides a problem formulation equivalent to (36).

**Lemma 8.** *Problem (36) is equivalent to*

$$\underset{\boldsymbol{\Omega}_i \succeq \mathbf{0}}{\text{maximize}} \; \log \det \left( \mathbf{I} + \boldsymbol{\Omega}_i \boldsymbol{\Sigma}_{\mathbf{y}_i | \hat{\mathbf{y}}_{\mathcal{N}_B \setminus \{i\}}} \right) - \log \det \left( \mathbf{I} + \boldsymbol{\Omega}_i \right) - q_H \text{tr} \left( \boldsymbol{\Omega}_i \right) \tag{54}$$

$$\text{s.t. } \log \det \left( \mathbf{I} + \boldsymbol{\Omega}_i \boldsymbol{\Sigma}_{\mathbf{y}_i | \hat{\mathbf{y}}_{\mathcal{N}_B \setminus \{i\}}} \right) \leq \bar{C}_i,$$

*where $\Sigma_{\tilde{\mathbf{y}}_i|\hat{\mathbf{y}}_{\mathcal{N}_{\mathcal{B}}\backslash\{i\}}}$ and $\bar{C}_i$ are defined in Theorem 2.*

*Proof:* We first observe that from the chain rule, the mutual information terms in problem (36) can be written as

$$I\left(\mathbf{x}; \hat{\mathbf{y}}_{\mathcal{S}_{\mathcal{H}}}|\hat{\mathbf{y}}_1\right) = I\left(\mathbf{x}; \hat{\mathbf{y}}_i|\hat{\mathbf{y}}_1, \hat{\mathbf{y}}_{\mathcal{S}_{\mathcal{H}}\backslash\{i\}}\right) + I\left(\mathbf{x}; \hat{\mathbf{y}}_{\mathcal{S}_{\mathcal{H}}\backslash\{i\}}|\hat{\mathbf{y}}_1\right), \tag{55}$$

and $I\left(\mathbf{y}_{\mathcal{S}_{\mathcal{H}}}; \hat{\mathbf{y}}_{\mathcal{S}_{\mathcal{H}}}|\hat{\mathbf{y}}_1\right) = I\left(\mathbf{y}_i; \hat{\mathbf{y}}_i|\hat{\mathbf{y}}_1, \hat{\mathbf{y}}_{\mathcal{S}_{\mathcal{H}}\backslash\{i\}}\right) + I\left(\mathbf{y}_{\mathcal{S}_{\mathcal{H}}\backslash\{i\}}; \hat{\mathbf{y}}_{\mathcal{S}_{\mathcal{H}}\backslash\{i\}}|\hat{\mathbf{y}}_1\right). \tag{56}$

In (55), the matrix $\mathbf{\Omega}_i$ affects only the first term which can be expressed as

$$I\left(\mathbf{x}; \hat{\mathbf{y}}_i|\hat{\mathbf{y}}_1, \hat{\mathbf{y}}_{\mathcal{S}_{\mathcal{H}}\backslash\{i\}}\right) = \log\det\left(\mathbf{I} + \mathbf{\Omega}_i\mathbf{\Sigma}_{\mathbf{y}_i|\hat{\mathbf{y}}_{\mathcal{N}_{\mathcal{B}}\backslash\{i\}}}\right) - \log\det\left(\mathbf{I} + \mathbf{\Omega}_i\right). \tag{57}$$

Thus, the objective function is given as (54). Similarly, the first and second terms in (56) are computed as

$$I\left(\mathbf{y}_i; \hat{\mathbf{y}}_i|\hat{\mathbf{y}}_1, \hat{\mathbf{y}}_{\mathcal{S}_{\mathcal{H}}\backslash\{i\}}\right) = \log\det\left(\mathbf{I} + \mathbf{\Omega}_i\mathbf{\Sigma}_{\mathbf{y}_i|\hat{\mathbf{y}}_{\mathcal{N}_{\mathcal{B}}\backslash\{i\}}}\right), \tag{58}$$

and $I\left(\mathbf{y}_{\mathcal{S}_{\mathcal{H}}\backslash\{i\}}; \hat{\mathbf{y}}_{\mathcal{S}_{\mathcal{H}}\backslash\{i\}}|\hat{\mathbf{y}}_1\right) = \log\det\mathbf{R}_i + \sum_{j\in\mathcal{S}_{\mathcal{H}}\backslash\{i\}}\log\det\left(\mathbf{I} + \mathbf{\Omega}_j\right), \tag{59}$

respectively, which results in the constraint of problem (54). ∎

Since problem (54) is non-convex, we first solve the KKT conditions, which can be proved to be necessary for optimality as in Lemma 4, and then show that the derived solution also satisfies the general sufficiency condition in [22, Proposition 3.3.4].

The Lagrangian of problem (54) is given as

$$\mathcal{L}\left(\mathbf{\Omega}_i, \mu, \mathbf{\Upsilon}\right) = (1-\mu)\log\det\left(\mathbf{I} + \mathbf{\Omega}_i\mathbf{\Sigma}_{\mathbf{y}_i|\hat{\mathbf{y}}_{\mathcal{N}_{\mathcal{B}}\backslash\{i\}}}\right) - \log\det\left(\mathbf{\Omega}_i + \mathbf{I}\right) - q_H\text{tr}\left(\mathbf{\Omega}_i\right) + \text{tr}\left(\mathbf{\Upsilon}\mathbf{\Omega}_i\right), \tag{60}$$

with Lagrange multipliers $\mu \geq 0$ and $\mathbf{\Upsilon} \succeq \mathbf{0}$. We then write the KKT conditions as

$$\frac{\partial\mathcal{L}}{\partial\mathbf{\Omega}_i} = \mathbf{0}, \tag{61a}$$

$$\text{tr}\left(\mathbf{\Upsilon}\mathbf{\Omega}_i\right) = 0, \tag{61b}$$

$$\mu\left(\log\det\left(\mathbf{I} + \mathbf{\Omega}_i\mathbf{\Sigma}_{\mathbf{y}_i|\hat{\mathbf{y}}_{\mathcal{N}_{\mathcal{B}}\backslash\{i\}}}\right) - \bar{C}_i\right) = 0, \tag{61c}$$

and $\log\det\left(\mathbf{I} + \mathbf{\Omega}_i\mathbf{\Sigma}_{\mathbf{y}_i|\hat{\mathbf{y}}_{\mathcal{N}_{\mathcal{B}}\backslash\{i\}}}\right) - \bar{C}_i \leq 0, \tag{61d}$

where the derivative $\partial\mathcal{L}/\partial\mathbf{\Omega}_i$ is given as

$$\frac{\partial\mathcal{L}}{\partial\mathbf{\Omega}_i} = (1-\mu)\mathbf{\Sigma}_{\mathbf{y}_i|\hat{\mathbf{y}}_{\mathcal{N}_{\mathcal{B}}\backslash\{i\}}}\left(\mathbf{I} + \mathbf{\Omega}_i\mathbf{\Sigma}_{\mathbf{y}_i|\hat{\mathbf{y}}_{\mathcal{N}_{\mathcal{B}}\backslash\{i\}}}\right)^{-1} - \left(\mathbf{\Omega}_i + \mathbf{I}\right)^{-1} - q'_H\mathbf{I} + \mathbf{\Upsilon}. \tag{62}$$

If we set $\Omega_i$ as (15) with the eigenvalue decomposition $\Sigma_{\tilde{\mathbf{y}}_i | \hat{\mathbf{y}}_{\mathcal{N}_\mathcal{B} \setminus \{i\}}} = \mathbf{U} \text{diag}(\lambda_1, \ldots, \lambda_{n_{B,i}}) \mathbf{U}^\dagger$, the KKT conditions (61a)-(61d) can be written as

$$\frac{(1-\mu)\lambda_l}{1+\alpha_l \lambda_l} - \frac{1}{1+\alpha_l} - q'_H + \theta_l = 0, \ \ l = 1, \ldots, n_{B,i} \tag{63a}$$

$$\theta_l \alpha_l = 0, \ \ l = 1, \ldots, n_{B,i} \tag{63b}$$

$$\mu \left( \sum_{l=1}^{n_{B,i}} \log \left(1 + \alpha_l \lambda_l\right) - \bar{C}_i \right) = 0, \tag{63c}$$

$$\text{and} \ \ \sum_{l=1}^{n_{B,i}} \log \left(1 + \alpha_l \lambda_l\right) - \bar{C}_i \le 0, \tag{63d}$$

with $\theta_l \ge 0$ for $l = 1, \ldots, n_{B,i}$. By direct calculation, we can see that the eigenvalues $\alpha_1, \ldots, \alpha_{n_{B,i}}$ in (38) satisfy the conditions (63a)-(63d).

Now, we show that the solution (38) also satisfies the general sufficiency condition in [22, Proposition 3.3.4].

**Lemma 9.** *Let $\Omega_i^*, \mu^*$ denote a pair obtained from Theorem 2. Then, $\Omega_i^*$ is the global optimum of problem (54) since it satisfies the sufficiency condition in [22, Proposition 3.3.4], i.e., the equality*

$$\Omega_i^* = \arg \max_{\Omega_i \succeq \mathbf{0}} \mathcal{L} \left( \Omega_i, \mu^* \right), \tag{64}$$

*and the complementary slackness condition (61c), where function $\mathcal{L} \left( \Omega_i, \mu \right)$ is the Lagrangian function in (60) with $\Upsilon = \mathbf{0}$.*

*Proof:* In problem (64), selecting the eigenvectors of $\Omega_i$ as those of $\Sigma_{\mathbf{y}_i | \hat{\mathbf{y}}_{\mathcal{N}_\mathcal{B} \setminus \{i\}}}$ does not involve any loss of optimality due to the eigenvalue inequality $\log \det \left( \mathbf{I} + \mathbf{AB} \right) \le \log \det \left( \mathbf{I} + \Gamma_\mathbf{A} \Gamma_\mathbf{B} \right)$ where $\Gamma_\mathbf{A}$ and $\Gamma_\mathbf{B}$ are diagonal matrices with diagonal elements of the decreasing eigenvalues of $\mathbf{A}$ and $\mathbf{B}$, respectively, with $\mathbf{A}, \mathbf{B} \succeq \mathbf{0}$ (see [11, Appendix B]). Then, (64) is equivalent to showing that the eigenvalues $\alpha_1^*, \ldots, \alpha_{n_{B,i}}^*$ in (38) satisfy

$$\alpha_l^* = \arg \max_{\alpha_l \ge 0} \left( (1-\mu^*) \log \left(1 + \alpha_l \lambda_l\right) - \log(1+\alpha_l) - q_H \alpha_l \right), \ \ \text{for } l = 1, \ldots, n_{B,i}. \tag{65}$$

The above condition can be shown by following the same steps as in [11, Appendix B]. ∎

## REFERENCES

[1] P. Marsch, B. Raaf, A. Szufarska, P. Mogensen, H. Guan, M. Farber, S. Redana, K. Pedersen and T. Kolding, "Future mobile communication networks: challenges in the design and operation," *IEEE Vehicular Technology Magazine*, vol. 7, no. 1, pp.16-23, March 2012.

[2] D. Gesbert, S. Hanly, H. Huang, S. Shamai (Shitz), O. Simeone and W. Yu, "Multi-cell MIMO cooperative networks: a new look at interference," *IEEE Jour. Sel. Areas Comm.*, vol. 28, no. 9, pp.1380-1480, Dec. 2010.

[3] A. Sanderovich, S. Shamai (Shitz), Y. Steinberg and G. Kramer, "Communication via decentralized processing," *IEEE Trans. Inf. Theory*, vol. 54, no. 7, pp. 3008-3023, July 2008.

[4] A. Sanderovich, S. Shamai (Shitz) and Y. Steinberg, "Distributed MIMO receiver-achievable rates and upper bounds," *IEEE Trans. Inf. Theory*, vol. 55, no. 10, pp. 4419-4438, Oct. 2009.

[5] A. Sanderovich, O. Somekh, H. V. Poor and S. Shamai (Shitz), "Uplink macro diversity of limited backhaul cellular network," *IEEE Trans. Inf. Theory*, vol. 55, no. 8, pp. 3457-3478, Aug. 2009.

[6] P. Marsch and G. Fettweis, "Uplink CoMP under a constrained backhaul and imperfect channel knowledge," *IEEE Trans. Wireless Comm.*, vol. 10, no. 6, pp. 1730-1742, June 2011.

[7] B. Nazer, A. Sanderovich, M. Gastpar and S. Shamai (Shitz), "Structured superposition for backhaul constrained cellular uplink," in *Proc. IEEE ISIT '09*, Seoul, Korea, June 2009.

[8] S.-N. Hong and G. Caire, "Quantized compute and forward: a low-complexity architecture for distributed antenna systems," in *Proc. IEEE ITW '11*, Paraty, Brazil, Oct. 2011.

[9] J. Chen and T. Berger, "Successive Wyner-Ziv coding scheme and its application to the quadratic Gaussian CEO problem," *IEEE Trans. Inf. Theory*, vol. 54, no. 4, pp. 1586-1603, April 2008.

[10] M. Gastpar, P. L. Dragotti and M. Vetterli, "The distributed Karhunen-Loeve transform," *IEEE Trans. Inf. Theory*, vol. 52, no. 12, pp. 5177-5196, Dec. 2006.

[11] A. del Coso and S. Simoens, "Distributed compression for MIMO coordinated networks with a backhaul constraint," *IEEE Trans. Wireless Comm.*, vol. 8, no. 9, pp. 4698-4709, Sep. 2009.

[12] G. Chechik, A. Globerson, N. Tishby and Y. Weiss, "Information bottleneck for Gaussian variables," *Jour. Machine Learn.*, Res. 6, pp. 165-188, 2005.

[13] A. Globerson and N. Tishby, "On the optimality of the Gaussian information bottleneck curve," *Hebrew University Technical Report*, 2004.

[14] C. Yu and G. Sharma, "Distributed estimation and coding: a sequential framework based on a side-informed decomposition," *IEEE Trans. Sig. Proc.*, vol. 59, no. 2, pp. 759-773, Feb. 2011.

[15] N. Tishby, F. C. Pereira and W. Bialek, "The information bottleneck method," in *Proc. 37th Allerton Conf.*, UIUC, pp. 368-377, Sep. 1999.

[16] A. D. Wyner and J. Ziv, "The rate-distortion function for source coding with side information at the decoder," *IEEE Trans. Inf. Theory*, vol. 22, no. 1, pp. 1-10, Jan. 1976.

[17] Y. C. Eldar, "Robust competitive estimation with signal and noise covariance uncertainties," *IEEE Trans. Inf. Theory*, vol. 52, no. 10, pp. 4532-4547, Oct. 2006.

[18] R. Mittelman and E. L. Miller, "Robust estimation of a random parameter in a Gaussian linear model with joint eigenvalue and elementwise covariance uncertainties," *IEEE Trans. Sig. Proc.*, vol. 58, no. 3, pp. 1001-1011, Mar. 2010.

[19] S Loyka and C. D. Charalambous, "On the compound capacity of a class of MIMO channels subject to normed uncertainty," *IEEE Trans. Inf. Theory*, vol. 58, no. 4, pp. 2048-2063, April 2012.

[20] A. Wiesel, Y. C. Eldar and S. Shamai (Shitz), "Optimization of the MIMO Compound Capacity," *IEEE Trans. Wireless Comm.*, vol. 6, no. 3, pp. 1094-1101, Mar. 2007.

[21] S. Boyd and L. Vandenberghe, Convex optimization, Cambridge University Press, 2004.

[22] D. Bertsekas, *Nonlinear programming*. New York: Athena Scientific, 1995.

[23] M. Hong, R.-Y. Sun and Z.-Q. Luo, "Joint base station clustering and beamformer design for partial coordinated transmission in heterogenous networks," arXiv:1203.6390.

[24] S. Ramanath, V. Kavitha and E. Altman, "Open loop optimal control of base station activation for green networks," in *Proc. WiOpt '11*, Princeton, NJ, May 2011.

[25] A. E. Gamal and Y.-H. Kim, *Network information theory*, Cambridge University Press, 2011.

[26] A. B. Wagner, S. Tavildar and P. Viswanath, "Rate region of the quadratic Gaussian two-encoder source-coding problem," *IEEE Trans. Inf. Theory*, vol. 54, no. 5, pp. 1938-1961, May 2008.

[27] I. Maric, B. Bostjancic and A. Goldsmith, "Resource allocation for constrained backhaul in picocell networks," in *Proc. ITA '11*, UCSD, Feb. 2011.

[28] T. M. Cover and J. A. Thomas, Elements of information theory. New York: Wiley, 2006.

[29] D. Bertsimas, D. B. Brown and C. Caramanis, "Theory and applications of robust optimization," arXiv:1010.5445.

[30] R. A. Horn and C. R. Johnson, *Topics in matrix analysis*, Cambridge University Press, 1991.

[31] C. T. K. Ng, C. Tian, A. J. Goldsmith and S. Shamai (Shitz), "Minimum expected distortion in Gaussian source coding with fading side information," arxiv.org/abs/0812.3709.

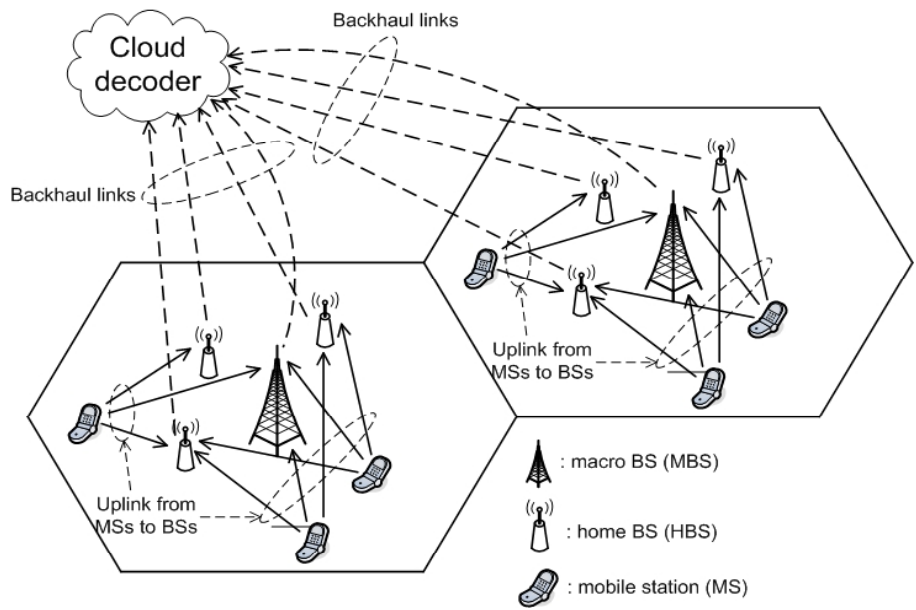

Figure 1. Uplink of a cloud radio access networks with BSs classified as Home BSs (HBSs) and Macro BSs (MBSs).

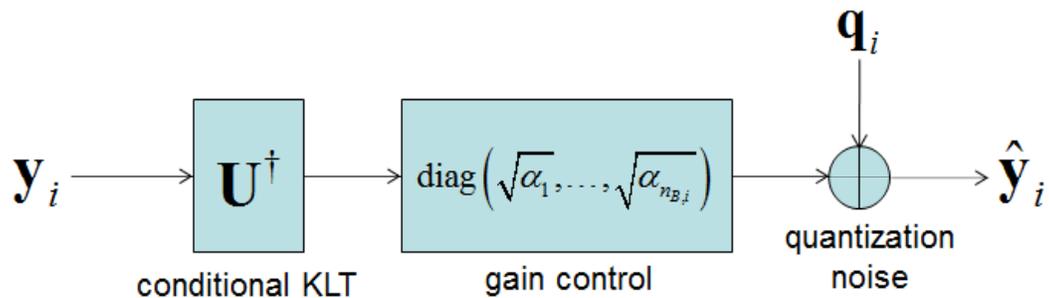

Figure 2. Illustration of the Max-Rate compression: $\mathbf{U}$ is the so called conditional KLT [10] and $\mathbf{q}_i$ represents the compression noise, with $\mathbf{q}_i \sim \mathcal{CN}(0, \mathbf{I})$.

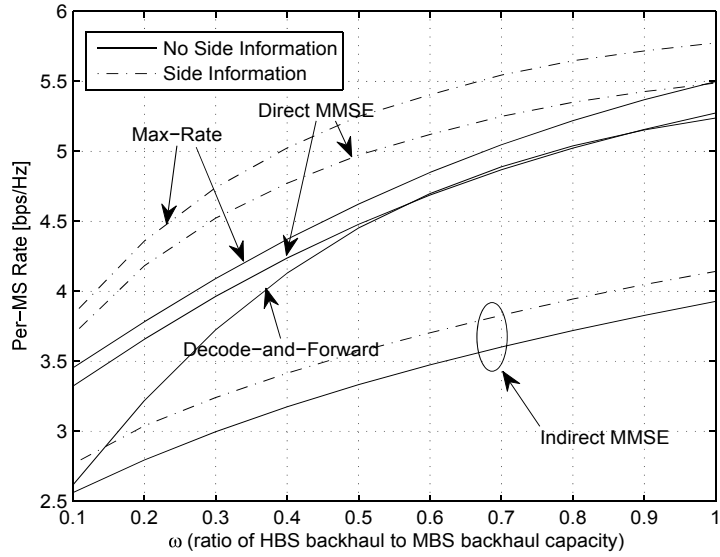

Figure 3.  Average per-MS sum-rate versus the ratio $\omega$ between the backhaul of the HBSs and of the MBS obtained with different compression methods for a three-cell heterogenous network with $N_B = 9$, $N_M = 9$, $n_{B,i} = 8$ and $C = 15$ bps/Hz at $P_{\mathrm{tx}} = -5$dB.

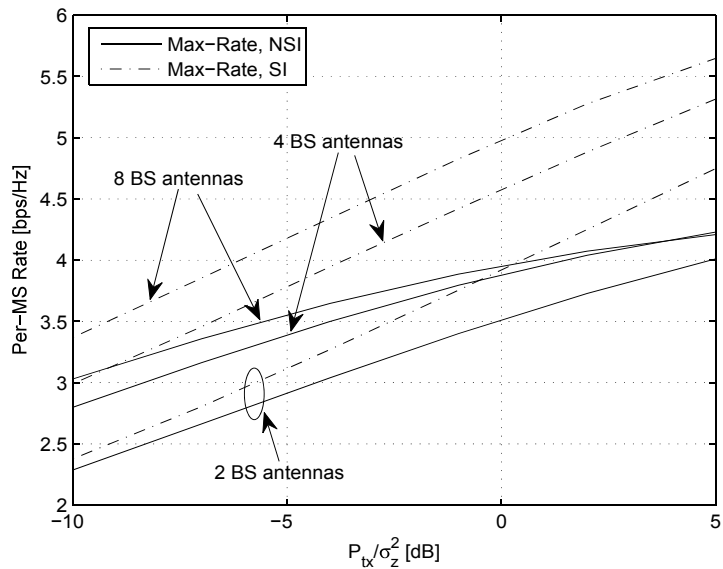

Figure 4.  Average per-MS sum-rate versus the SNR $P_{\mathrm{tx}}$ obtained with the Max-Rate compression for a three-cell heterogeneous network with $N_B = 9$, $N_M = 9$, $\omega = 0.5$ and $C = 10$ bps/Hz.

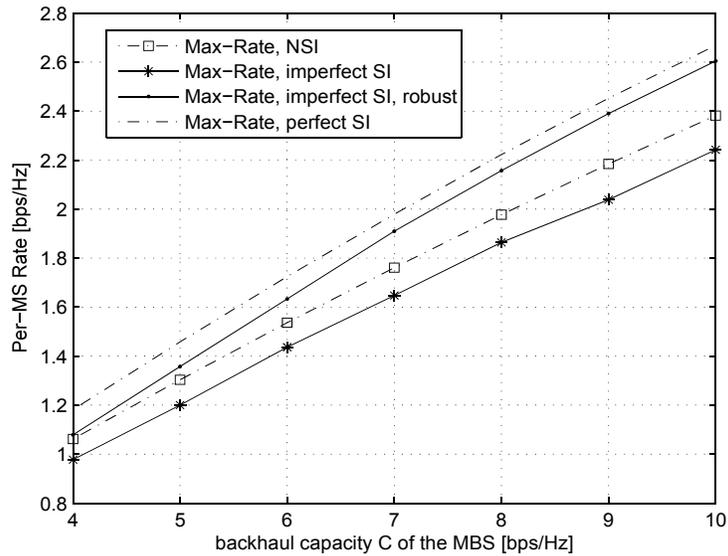

Figure 5. Average per-MS sum-rate versus the backhaul capacity $C$ the MBS obtained with the Max-Rate compression scheme in the presence of uncertainty for a single-cell heterogeneous network with $N_B = 4$, $N_M = 8$, $n_{B,i} = 2$ and $\omega = 0.5$ at $P_{\mathrm{tx}} = 10$dB.

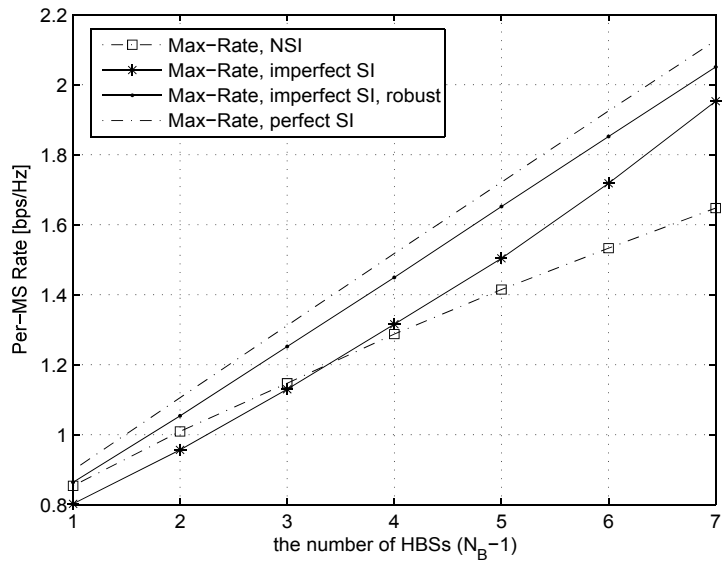

Figure 6. Average per-MS sum-rate versus the number $N_B - 1$ of HBSs obtained with the Max-Rate compression scheme in the presence of uncertainty for a single-cell heterogeneous network with $N_M = 10$, $n_{B,i} = 8$ and $C = 7$ bps/Hz at $P_{\mathrm{tx}} = 0$dB.

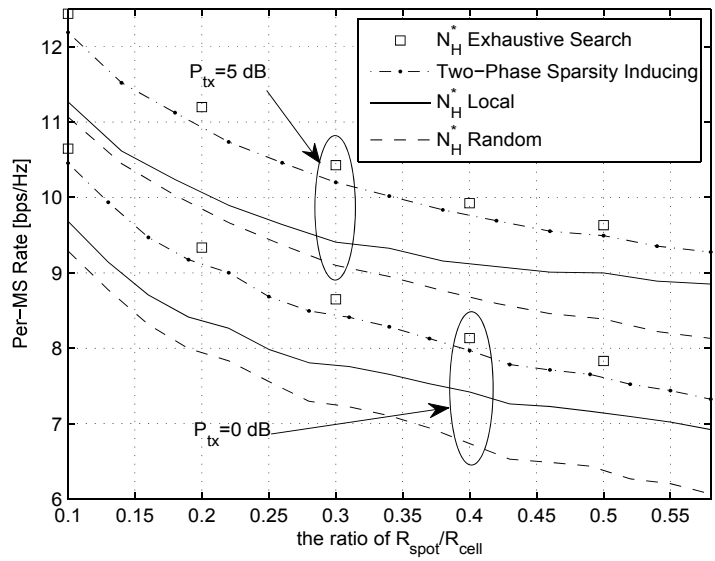

Figure 7. Average per-MS sum-rate versus the ratio $R_{\text{spot}}/R_{\text{cell}}$ in a single-cell heterogeneous network with $N_B = 13$ ($N_B^1 = 6$, $N_B^2 = 6$), $N_M = 14$ ($N_M^1 = 8$, $N_M^2 = 6$), $n_{B,i} = 8$, $C = 20$ bps/Hz, $C_H = 300$ bps/Hz and $q_H = 100$.